\documentclass[12pt]{article}\usepackage[hyperfootnotes=false]{hyperref}
\usepackage{epsfig}
\usepackage{float}

\usepackage{caption}

\usepackage{amsmath}
\usepackage{amssymb}
\usepackage{graphicx}
\newlength{\abstractwidth}
\renewcommand{\thefootnote}{\fnsymbol{footnote}}
\renewcommand{\thanks}[1]{\footnote{#1}} 
\newcommand{\starttext}{
\setcounter{footnote}{0}
\renewcommand{\thefootnote}{\arabic{footnote}}}

\newcommand{\be}{\begin{equation}}
\newcommand{\bea}{\begin{eqnarray}}
\newcommand{\eea}{\end{eqnarray}}
\newcommand{\beq}{\begin{equation}}
\newcommand{\ee}{\end{equation}}

\def\eq{&=&}

\def\ra{\rangle}

\def\simleq{\; \raise0.3ex\hbox{$<$\kern-0.75em
\raise-1.1ex\hbox{$\sim$}}\; }
\def\simgeq{\; \raise0.3ex\hbox{$>$\kern-0.75em
\raise-1.1ex\hbox{$\sim$}}\; }

\def\bi{\begin{itemize}}
\def\ei{\end{itemize}}

\def\sc{\setcounter{equation}{0}}

\def\CI{Copenhagen Interpretation }
\def\RSF{Relative State Formulation }
\def\WF{Wigner's Friend }

\def\bn{\bigskip \noindent}

\def\erd{Einstein-Rosen bridge. }

\def\bn{\bigskip \noindent}

\makeatletter
\g@addto@macro\normalsize{%
  \setlength\abovedisplayskip{10pt}
  \setlength\belowdisplayskip{20pt}
  \setlength\abovedisplayshortskip{10pt}
  \setlength\belowdisplayshortskip{20pt}
}
\makeatother

\usepackage{color}

\begin{document}
  
\begin{titlepage}

\rightline{}
\bigskip
\bigskip\bigskip\bigskip\bigskip
\bigskip

\centerline{\Large \bf {Copenhagen vs Everett, }}

\bn

\centerline{\Large \bf {Teleportation,  and ER=EPR }}

\bigskip
\begin{center}
\bf   Leonard Susskind  \rm

\bigskip

 Stanford Institute for Theoretical Physics and Department of Physics, \\
Stanford University,
Stanford, CA 94305-4060, USA \\
\bigskip

\end{center}

\begin{abstract}
Quantum gravity may have as much to tell us about the foundations and interpretation of quantum mechanics as it does about gravity. 
The Copenhagen interpretation of quantum mechanics and  Everett's Relative State Formulation are complementary descriptions which in a sense are dual to one another. My purpose here is to discuss this duality in the light of the  of  ER=EPR conjecture.

\medskip
\noindent
\end{abstract}

\end{titlepage}

\starttext \baselineskip=17.63pt \setcounter{footnote}{0}
Let me start with three quotes from famous quantum physicists.

\bn

Niels Bohr:

\it
If quantum mechanics hasn't profoundly shocked you, you haven't understood it yet\footnote{To Everett it may have seemed that Bohr was not sufficiently shocked.}.   

 \rm

\bn 

Richard Feynman:

\it
We have always had a great deal of difficulty understanding the world view that quantum mechanics represents. At least I do, because I'm an old enough man that I haven't got to the point that this stuff is obvious to me. Okay, I still get nervous with it.... You know how it always is, every new idea, it takes a generation or two until it becomes obvious that there's no real problem. I cannot define the real problem, therefore I suspect there's no real problem, but I'm not sure there's no real problem.

\rm

\bn

Paul Dirac:
\it

There is hope that quantum mechanics will gradually lose its baffling quality...... I have observed in teaching quantum mechanics, and also in learning it, that students go through an experience.... The student begins by learning the tricks of the trade. He learns how to make calculations in quantum mechanics and get the right answers.....it is comparatively painless. The second stage comes when the student begins to worry because he does not understand what he has been doing. He worries because he has no clear physical picture in his head..... Then, unexpectedly, the third stage begins. The student suddenly says to himself, “I understand quantum mechanics,” or rather he says, “I understand now that there isn't anything to be understood.”.... The duration and severity of the second stage are decreasing as the years go by. Each new generation of students learns quantum mechanics more easily\footnote{Too easily? Sometimes I think so.} than their teachers learned it..... \rm

\bn
 Is there a problem, as Feynman suggests---and then suggests there isn't---and then suggests that maybe there is?  What is it that has caused so much angst and spilled ink?  Pretty clearly it's about
the confusing relation between the multiplicity of observers,  and the objects
of their observations; namely each other and the rest of the universe. Standard quantum mechanics---
Copenhagen quantum mechanics---is set up in a way that requires a single external observer, one 
who is not part of the system. He, she, or it interacts occasionally with the system through
a process called measurement, and in so doing 
 collapses the wave function,  throwing away all  branches other than the one observed.  In this view observations are irreversible events which cannot be undone.
The method works well in  practice but only because reversing a measurement is generally too complex a process to ever be of practical importance.

It is obvious that the Copenhagen Interpretation  cannot be the last word. The universe is filled with subsystems, any one of which can play the role of observer. There is no place  in the laws of quantum mechanics for wave function collapse; the only thing that happens is that the overall wave function evolves unitarily and becomes more and more entangled. The universe is an immensely complicated network of entangled subsystems, and only in some approximation can we single out a particular subsystem as THE OBSERVER. 
\bn

 Until recently my view about these things  was pretty similar to Feynman's with maybe a bit of Dirac's; namely: \it Quantum mechanics is so confusing that I can't even tell if there is a problem, but maybe it's all ok because it works. There is probably not much profit in thinking about ``interpretations" and even less in arguing about them. \rm
 
 \bn

But over the last two years I've come to see it differently. Now I feel that our current views of quantum mechanics are provisional; it's the best we can do without a much deeper understanding of its connection with gravity, but it's not final.  The reason  involves a very particular development, the so called ER=EPR principle. ER=EPR tells us that the \it immensely complicated network of entangled subsystems that comprises the universe \rm is also an immensely complicated (and technically complex) network of Einstein-Rosen bridges.
To me it seems obvious  that if ER=EPR is true it is a very big deal, and it must affect the foundations and interpretation of  quantum mechanics.

\bn

What follows is a lecture that I gave at the Institute for Advanced Study in March 2016. It's purpose was to point out some of the likely conceptual implications of ER=EPR. The lecture  consisted of three short seminars:

 1. ``ER=EPR: Everett vs Copenhagen" or ``GHZ-Branes"

 2.``Teleportation Through the Wormhole: ERBs as a Resource"

 3.``Two Slits and a Wormhole"

 If there is a common theme it is that recent developments in quantum gravity have as much to tell us about  quantum mechanics as they do about gravity.
The figures in the paper are  mostly the original slides for the lecture, and the text between the figures is what I explained verbally.

\sc
\section{ER=EPR: Everett vs Copenhagen: ``GHZ-Branes"}

\subsection*{The Meaning of ER=EPR}
Quantum mechanics requires a kind of non-locality called Einstein-Podolsky-Rosen  entanglement. EPR does not violate causality, but it is, nevertheless,  a form of non-locality. It is most clearly seen if one imagines trying to simulate quantum mechanics on a system of classical computers. We assume the computers are distributed throughout space and represent local degrees of freedom. The whole conglomeration is required to  behave as if there were quantum systems inside the computers; systems that local observers can ``observe" by pushing buttons and reading outputs. The computers will of course have to interact with each other, as they also would  if we were simulating classical physics. But  simulating classical  physics only requires the computers to interact with their local neighbors.

Simulating quantum mechanics is different. One finds that it is necessary to have all the computers wired to an enormous central memory in order to store entangled states\footnote{In this paper the term ``state" will only be used for pure states. }, and also a central random number generator to provide indeterminacy. 
\begin{figure}[H]
\begin{center}
\includegraphics[scale=.3]{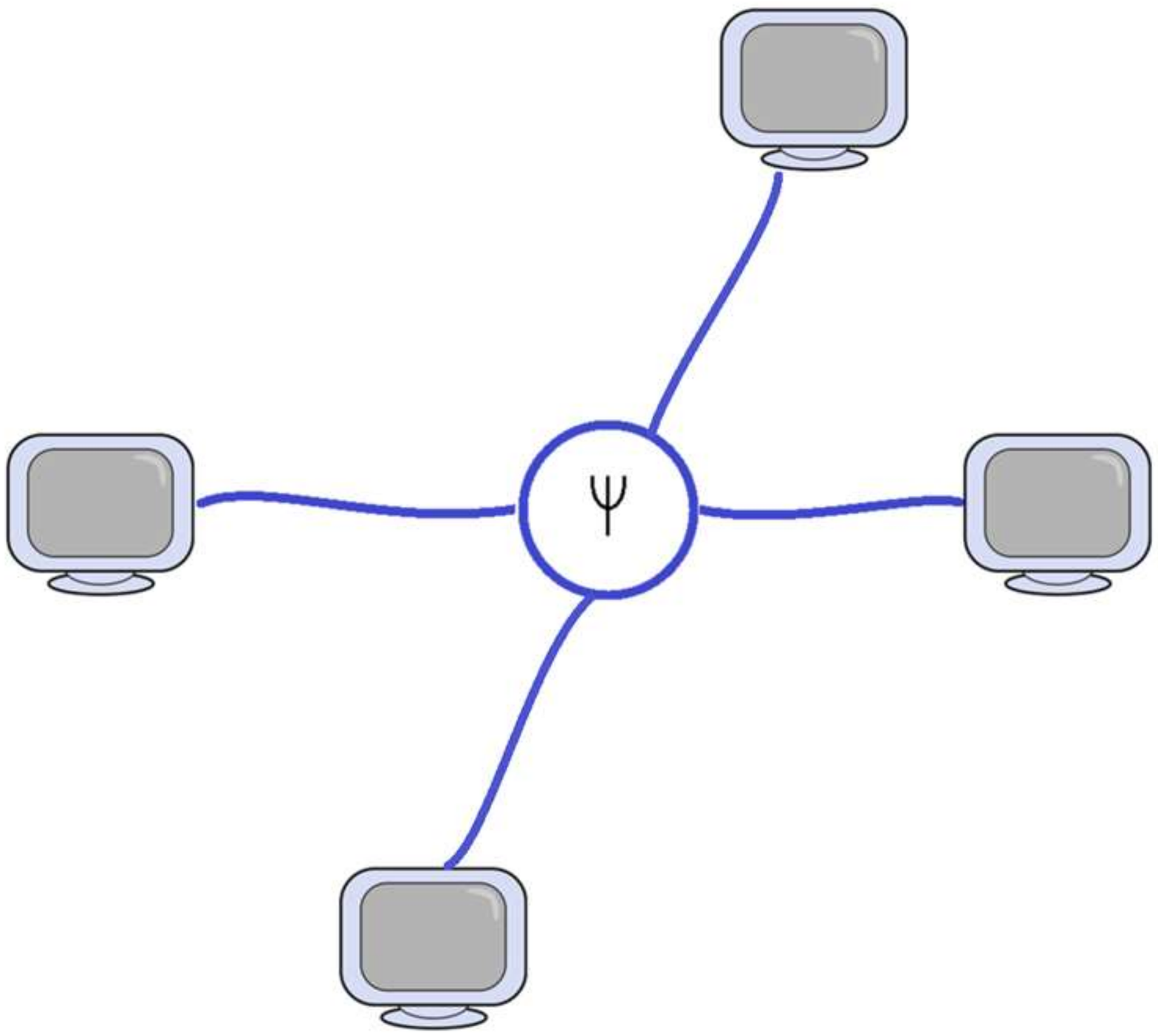}
\caption{}
\label{f1}
\end{center}
\end{figure}
Moreover the wires which connect the classical computers to the central memory must be able to  transmit signals instantaneously. Of course this is not to say that quantum mechanics allows instantaneous signaling; only that in  order to simulate QM on classical machines there must be instantaneous transmission. Figure \ref{f1} is a cartoon of several remote computers connected to a memory that stores entangled states.

\bn

In the next figure space is shown ``folded" in order to draw spatially distant points close to one another. 
\begin{figure}[H]
\begin{center}
\includegraphics[scale=.5]{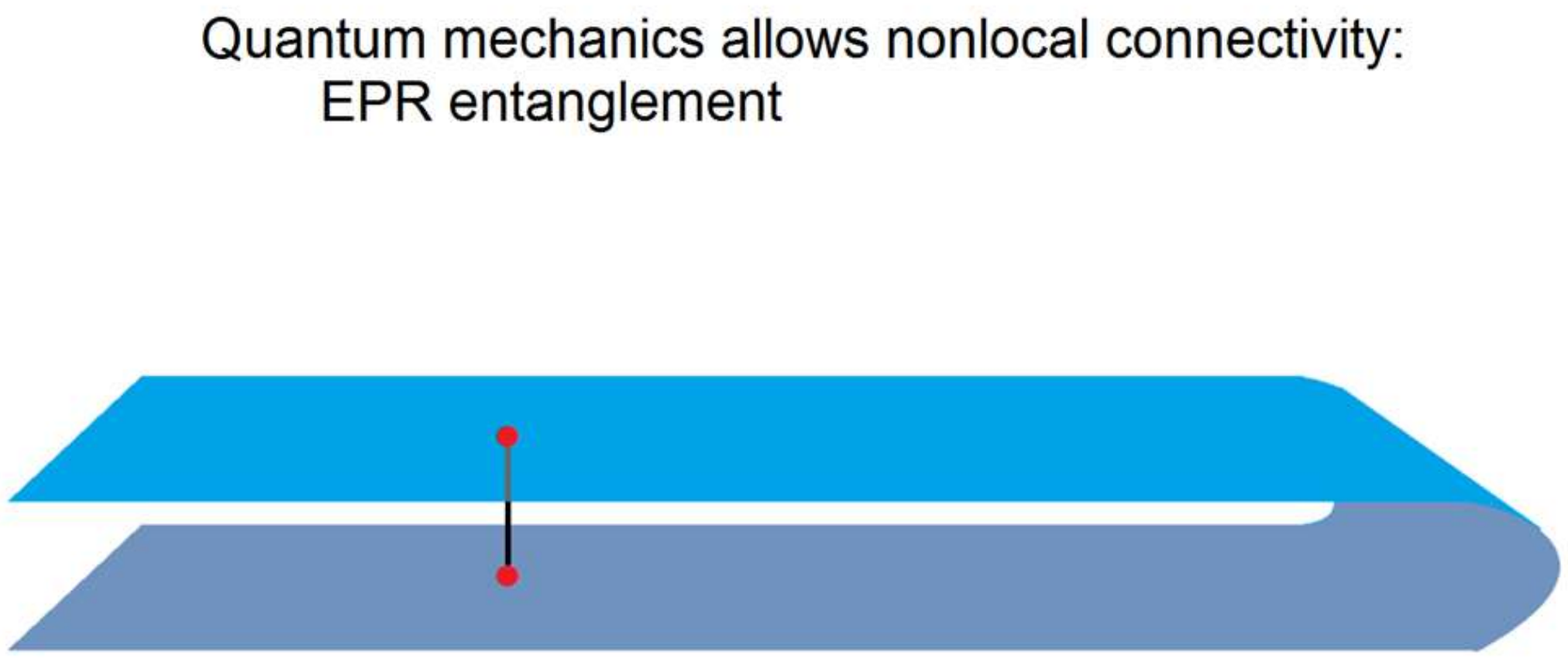}
\caption{}
\label{f2}
\end{center}
\end{figure}
\bn
The two red dots are maximally entangled particles and I indicate their entanglement by linking them by a short black line. The black link has some structure; for example it distinguishes between the various maximally entangled Bell states.
Despite appearances the nonlocal features of entanglement cannot be used to transmit messages superluminally (faster than light). 

\bn

General Relativity also has its non-local features. In particular there are solutions to Einstein's equations in which a pair of arbitrarily distant black holes are connected by a wormhole or Einstein-Rosen bridge (ERB).
\begin{figure}[H]
\begin{center}
\includegraphics[scale=.5]{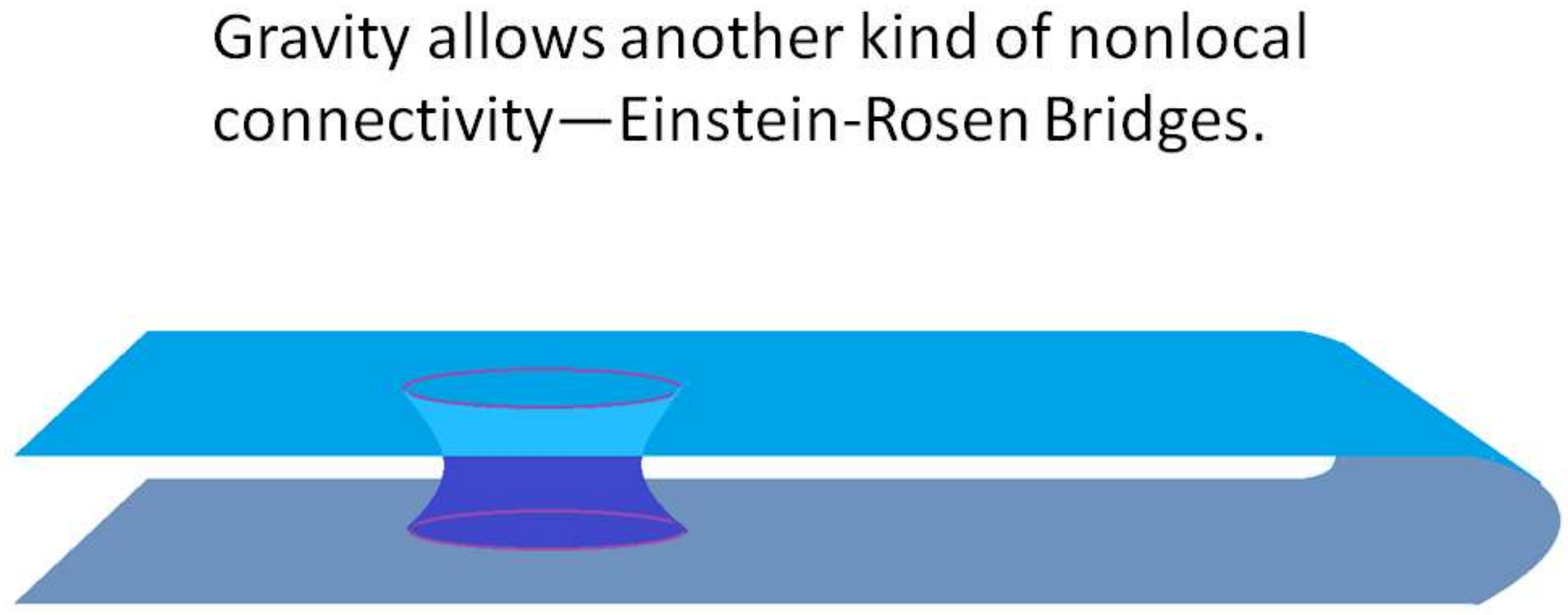}
\caption{}
\label{f3}
\end{center}
\end{figure}
\bn
 At first sight it would seem that  ERBs can  be used to superluminally transmit  signals. But this is not so; the wormhole solutions of general relativity are ``non-traversible."  (Non-traversibility means that two observers just outside the  black holes cannot communicate through the ERB. Non-traversibility does allow them to jump in and meet in the ERB.)
 
 The similarity between figures \ref{f2} and \ref{f3} is  quite intentional.
The punchline of the ER=EPR joke is that in some sense the phenomena of Einstein-Rosen bridges and Einstein-Podolsky-Rosen entanglement are really   the same \cite{Maldacena:2013xja}:

\bn

\ \ \ \ \ \ \ \ \ \ \ \ \ \ \ \ \ \ {\large{\bf ER=EPR.}}

\bn

This is a remarkable claim whose impact has yet to be appreciated.
There are two views of what it means, one modest and one more ambitious. The ambitious view is that some future conception of quantum geometry will even allow us to think of  two entangled spins---a Bell pair---as being connected by a Planckian wormhole. For most of this lecture the more modest definition will suffice, but toward the end I'll speculate about the ambitious view.

The modest view first of all says that black holes connected by ERBs are entangled and also the converse;  entangled black holes are connected by ERBs. But there is more to it than that. The idea can be stated in terms of entanglement being  a ``fungible resource." 
Entanglement is a resource because it is useful for carrying out certain communication tasks such as teleportation  \cite{Bennett}\cite{Wooters}\cite{HHHH}\cite{Elbowgrease}. It is 
fungible because like energy, which comes in different forms---electrical, mechanical, chemical, etc.---entanglement also comes in many forms  which can be transformed into one another. Some forms of entanglement:

\bn
1. Ground state or vacuum entanglement;

\bn
2. Entangled particles;

\bn
3. Einstein-Rosen bridges; 

\bn
4. See last part of this three-part lecture.

\bn

What about the conservation of the resource?
Energy is conserved but entanglement is not, except under special circumstances. If two systems are distantly separated  so that they can't interact, then the entanglement between them is conserved under independent local unitary transformations. Thus if Alice and Bob, who are far from one another, are each in control of two halves of an entangled system,  the unitary manipulations they do on their own shares cannot change the entanglement entropy.

If   Alice's  system interacts with a nearby environment, the entanglement with Bob's system can be transferred to the environment, but as long as the environment stays on Alice's side and does not interact with Bob's system the entanglement will be conserved.

\begin{figure}[H]
\begin{center}
\includegraphics[scale=.5]{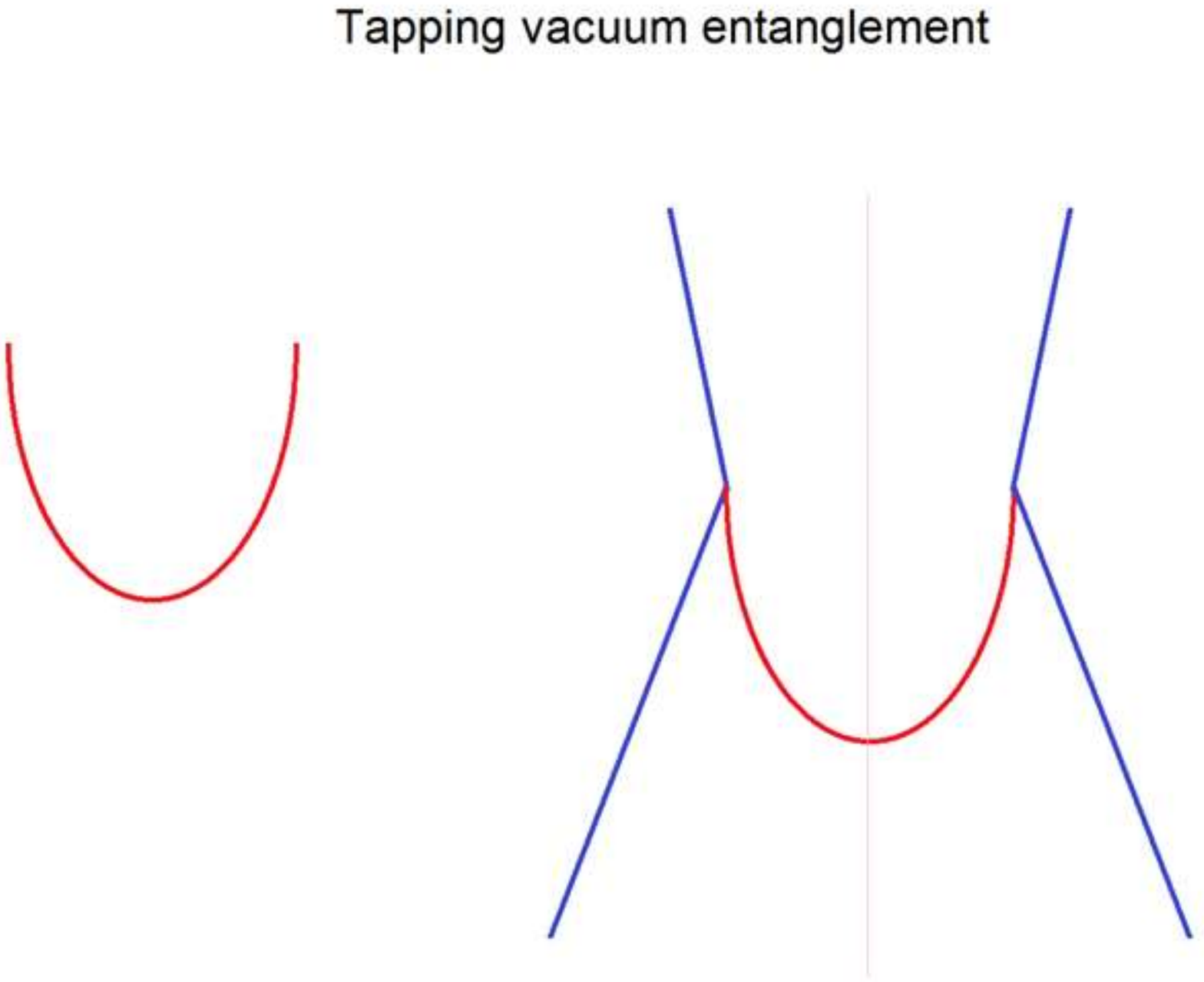}
\caption{}
\label{f6}
\end{center}
\end{figure}

Let's consider some examples of the transformation of entanglement from one form to another. In the  vacuum of a quantum field theory the quantum fields in disjoint regions of space are entangled.
One way to picture  this  is that virtual pairs of entangled particles are constantly appearing for short times, as  in the left side of figure \ref{f6}. 
\bn
If, as in the right side of the figure, two unentangled real particles  scatter off the virtual particles, they can leave in an entangled Bell state.

Now let's suppose that after making a large number of such Bell pairs, Alice takes half of each pair and Bob takes the other half. They separate to a large distance, maintaining the entanglement resource.
\begin{figure}[H]
\begin{center}
\includegraphics[scale=.3]{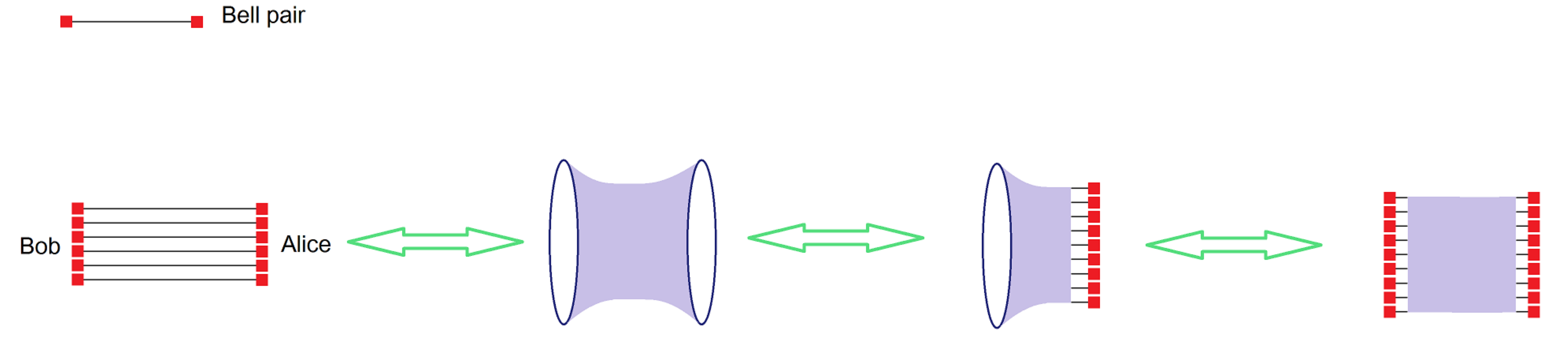}
\caption{}
\label{f7}
\end{center}
\end{figure} 
\bn
Then each   takes  their cloud of particles and compresses it to form a black hole. By ER=EPR the two black holes  will be connected by  a wormhole. By this means  the  original entanglement resource may be converted  into an \erd

Alice may allow her black hole to evaporate, while  Bob keeps his  in a sealed reflecting box.   After a while Bob's black hole will be become entangled with Alice's cloud of particles.

If  Bob subsequently lets his black hole evaporate the result will be  an entangled system of particles which will be so completely scrambled that there will be a high degree of entanglement, no matter how  the system is partitioned. With  sufficiently powerful quantum computers Alice and Bob  can locally act on their shares  to covert them back to Bell pairs. 

In an information theoretic sense all these systems---Bell pairs, ERB's, and clouds of Hawking radiation---are equivalent, and can be transformed into each other \cite{Czech:2014tva}  by local unitary transformation, acting independently at Alice's and Bob's ends. While in practice applying such unitary operations may be impractical, we will assume that they are in-principle possible. This fungibility  allows us to reinterpret quantum phenomena, involving ordinary forms of  entanglement, in terms of the geometric properties of ERBs.

That's the modest meaning of ER=EPR. The more adventurous meaning is that even the simple Bell pair has a highly quantum version of 
an ERB connecting it, and when brought together with a great many other quantum ERBs, they merge to form a large ERB.

\sc
\subsection*{  Copenhagen vs Everett}

The evolution of a  quantum state is often described as a phase-coherent branching  tree of possibilities. In the Copenhagen Interpretation  the observer is considered to be outside the system, but can interact with it from time to time by the process of measurement.
\begin{figure}[H]
\begin{center}
\includegraphics[scale=.2]{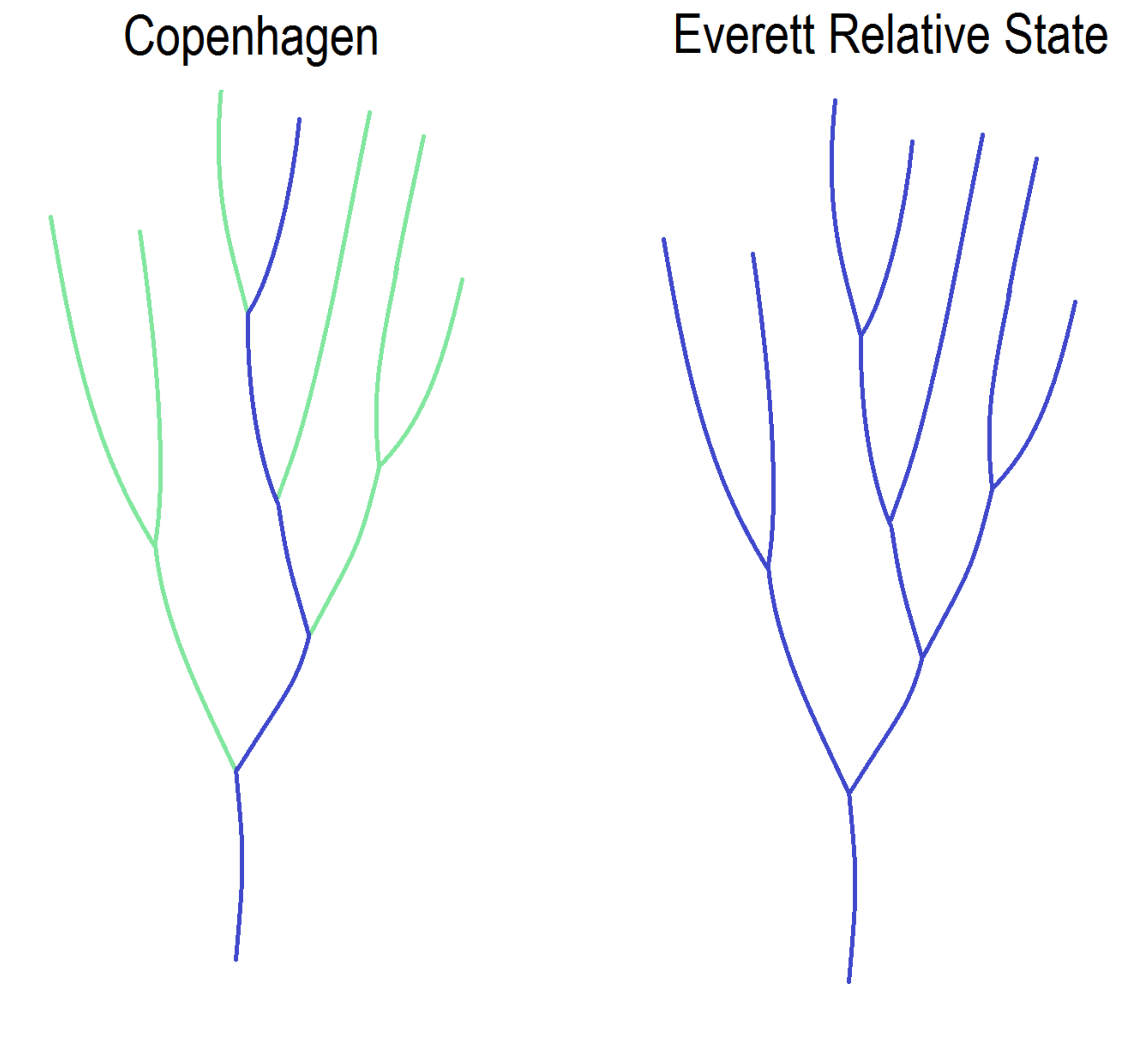}
\caption{}
\label{f9}
\end{center}
\end{figure}
\bn
Each measurement causes branches to decohere; the quantum superposition being replaced by classical probabilities. The observer follows a trajectory through the tree.
According to the \CI  \ measurements are irreversible, and the relative phases between branches lose all meaning. 

In practice this  is the standard working interpretation of quantum mechanics,  but it is not entirely consistent.
Everett's Relative-State Formulation  is an attempt to remedy the inconsistencies of the \CI
\footnote{Everett did not use the term ``many worlds," and didn't  like the idea. The metaphysical  interpretation was attributed to his work by Wheeler, DeWitt, and Graham. 
Here is a quotation from the Stanford Encyclopedia of Philosophy that can be found on the internet:

\it `` ....there was no consensus between Everett, Wheeler, DeWitt, and Graham concerning what Everett's theory was. In particular, we know what Everett thought of Graham's formulation of the theory. In his personal copy of DeWitt's description of the many worlds interpretation, Everett wrote the word `bullshit' next to the passage where DeWitt presented Graham's clarification of Everett's views." \rm 

I believe that  Everett's ideas are captured by the analysis of the Wigner's Friend story that follows.}.

According to Everett's Relative-State Formulation  there is only one system, the universe; all observers are part of it, and are subject to the unitary laws of quantum mechanics. The system can be divided into subsystems  any of which  can be regarded as an observer.  Collapse of the wave function never takes place; instead interactions cause subsystems to become entangled. The entire tree, i.e., the entire wave function,  must be retained and the universe is the complicated  network of entanglements that I referred to earlier.

In fact the tree is not a tree \cite{Bousso:2011up}. In principle everything is reversible. For example if the system is closed it will  execute quantum recurrences, eventually growing together and reforming the initial root. As we will see there are other faster ways  to reverse observations. Accordingly the relative phases in the wave function are always important and may not be thrown away.

The thing that in practice allows us to get away with the irreversible collapse postulate is the enormous technical complexity of ordinary observations which usually  makes it unfeasible to reverse them.

\subsection*{Wigner's Friend}
To illustrate the two ways of thinking we can consider a variant of the Schrodinger's Cat experiment, the ``Wigner's Friend Experiment." 
\begin{figure}[H]
\begin{center}
\includegraphics[scale=.3]{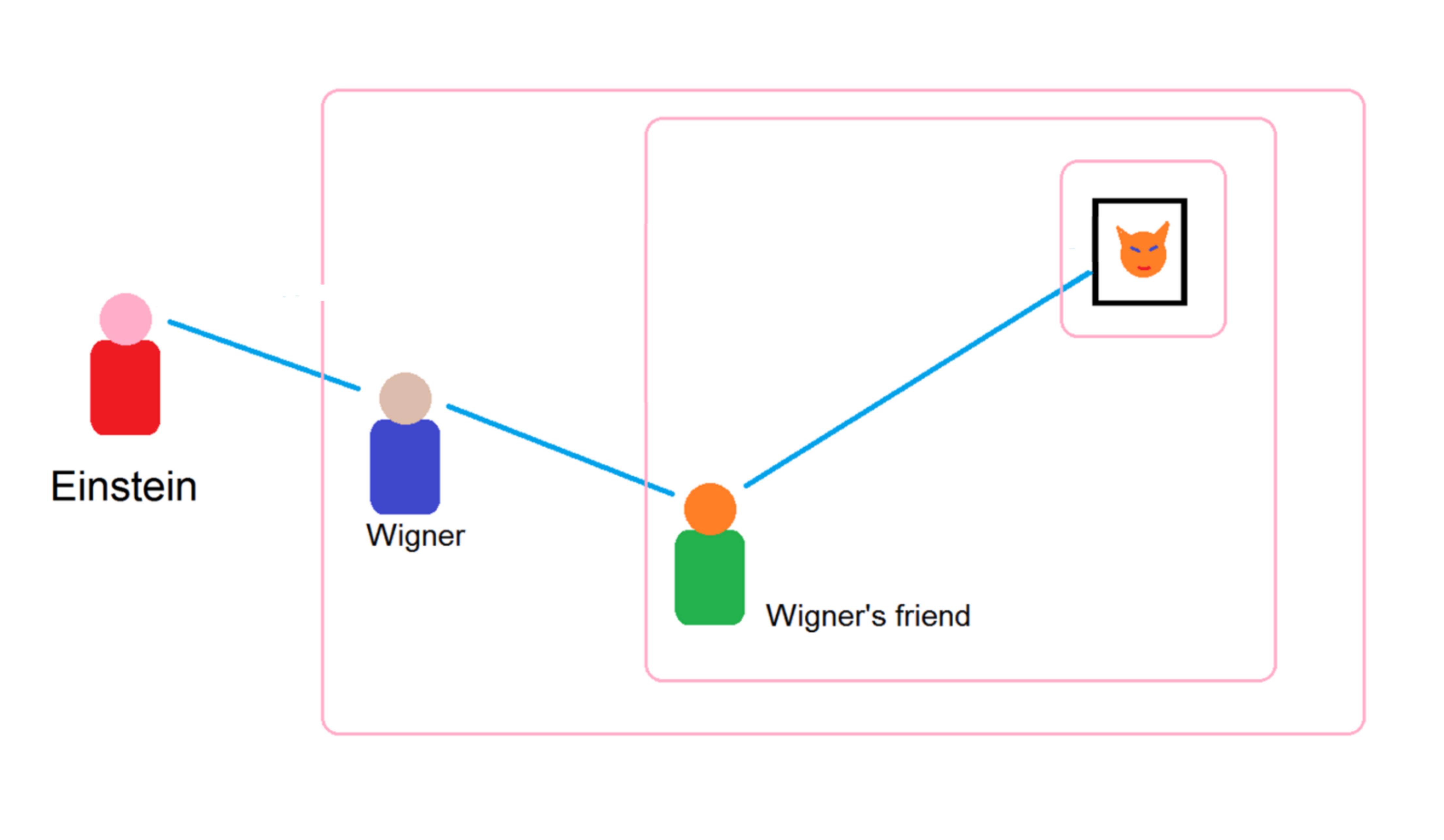}
\caption{}
\label{f10}
\end{center}
\end{figure}
\bn 
The original thought experiment involved two observers: Wigner and his friend. We'll add one more called Einstein\footnote{Why Einstein? I first used Schrodinger, but that so scared the cat  that she hid under the bed.}.

The experiment begins with a cat in a sealed room, and all three observers outside the room.  Initially the cat is in the usual cat-state, a superposition of dead and alive. Then
Wigner's Friend enters the room and observes the cat. The cat is of course a collection of $N$ qubits and for our purposes the observation is a measurement not only of whether the cat is dead or alive---that's just one qubit---but of all $N$  qubits. To be precise Wigner's Friend measures the $Z$ component of all $N$ qubits, i.e., all $N$  ``Z-bits."  

 In order to record the result Wigner's Friend must have a register or memory in which he classically records the value of every  Z-bit. It could  be an electronic memory, or just  a portion of  Wigner's Friend's brain.
According to the \CI the state of the cat collapses to one of $2^N$ states, and \WF  \ records the result in his register. The \CI assumes that there is no possibility of undoing the measurement. 

Now let's turn to the \RSF \  of the same event.   Wigner might use such a description before he enters the room, especially if he regards his friend and the cat as comprising a single quantum system. In the \RSF \ the interaction of Wigner's Friend with the cat does not cause a wave function collapse;
instead,  the qubits of Wigner's Friend's memory become maximally entangled with the cat's qubits.

\begin{figure}[H]
\begin{center}
\includegraphics[scale=.5]{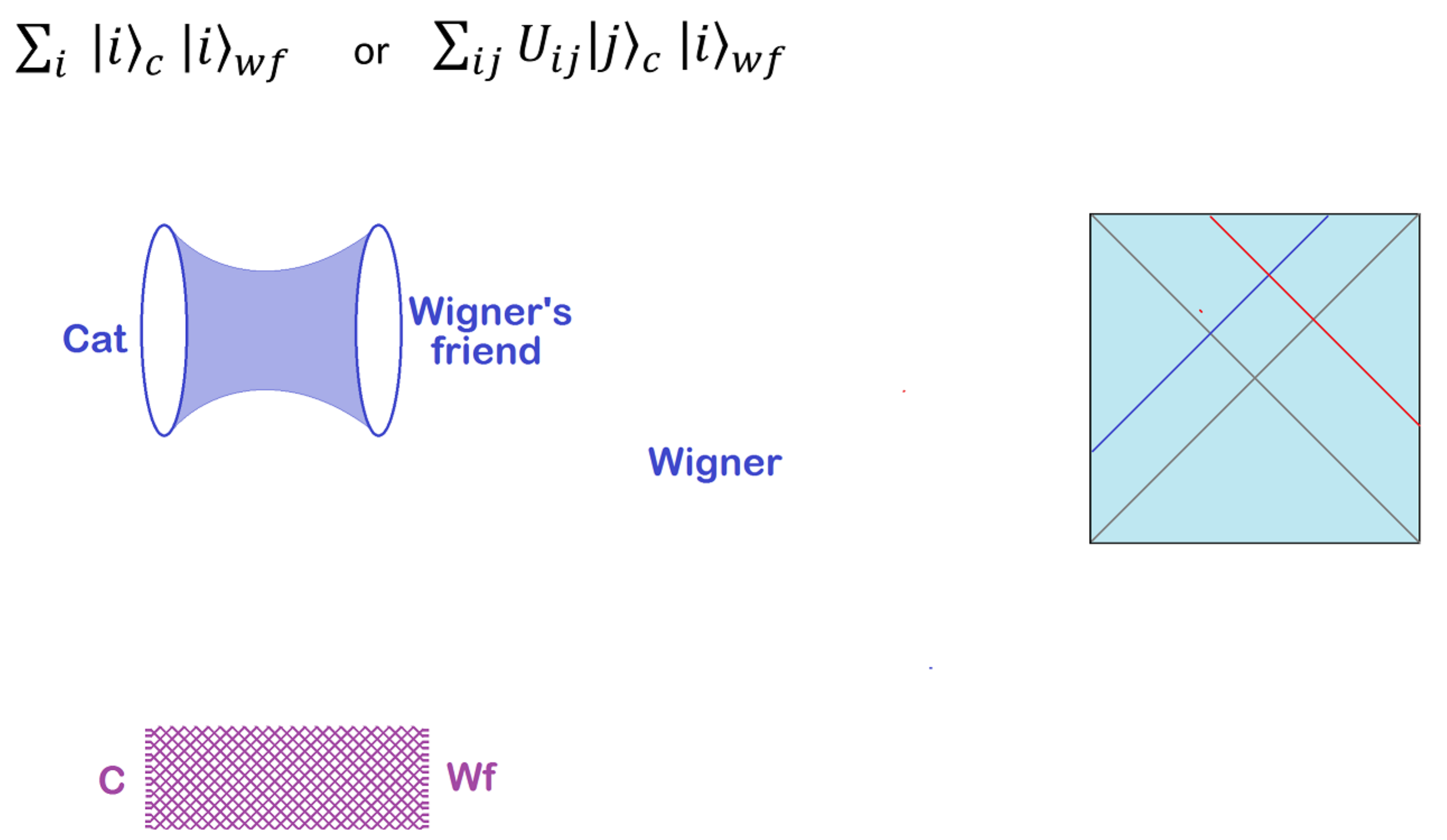}
\caption{}
\label{f11}
\end{center}
\end{figure}

If we believe in the ambitious form of ER=EPR we might say that \WF \ and the cat  become connected by some collection of quantum wormholes. But we can also appeal to the modest view; fungibility of entanglement allows us to 
compress both \WF \ and the cat into two black holes.
 The black holes will be entangled, and by virtue of that entanglement there will be an  ERB connecting them as in figure \ref{f11}.
\bn
As time evolves after the compression, the ERB will grow, as layers are added to the tensor network. The evolution is represented by a unitary matrix $U_{ij}$ and by a growing tensor network in figure \ref{f11}.
Following \cite{Maldacena:2013xja} we assume  that a sufficiently complex operation would allow messages (or observers) sent from the exteriors of the black holes  to meet inside the ERB.

Now let Wigner enter the room and observe his friend, again in the Z basis. 
In the \CI the observation irreversibly collapses the entangled state to a  single unentangled  product state. One can visualize this as a ``snipping"   process which cuts  the ERB at  Wigner's Friend's end \cite{Susskind:2014yaa}. 
\begin{figure}[H]
\begin{center}
\includegraphics[scale=.25]{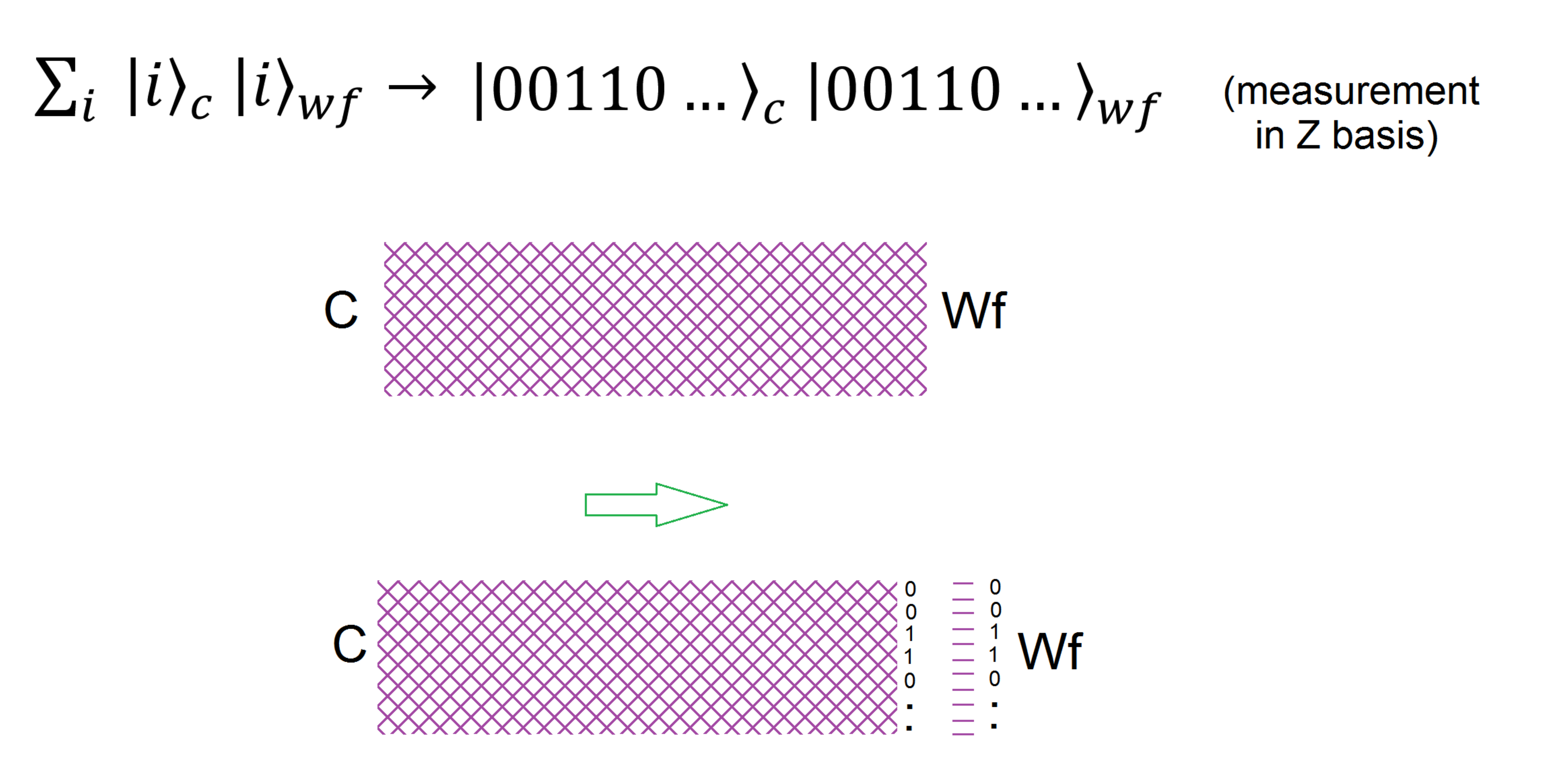}
\caption{}
\label{f12}
\end{center}
\end{figure}
\bn
If any appreciable evolution of the ERB had taken place (growth of the tensor network) the cat would be left in a somewhat complex state, but Wigner's Friend, having been projected onto a state $|00110...\ra,$ would be in a simple unentangled state.
Evidently, since according to the \CI the cat and \WF are no longer entangled, messages from the two ends cannot meet in the wormhole.

After the ERB is snipped the complexity of the two black holes will grow \cite{Susskind:2014rva}\cite{Stanford:2014jda}, and so will  the volumes of the two 
``bridges-to-nowhere\footnote{One-sided black holes in pure states do not have bridges connecting them to other systems, but they do have growing interiors which resemble bridges-to-nowhere \cite{Susskind:2014jwa}.}." But the bridge between the cat and Wigner's Friend will remain cut as long as the two systems don't directly interact.
\begin{figure}[H]
\begin{center}
\includegraphics[scale=.3]{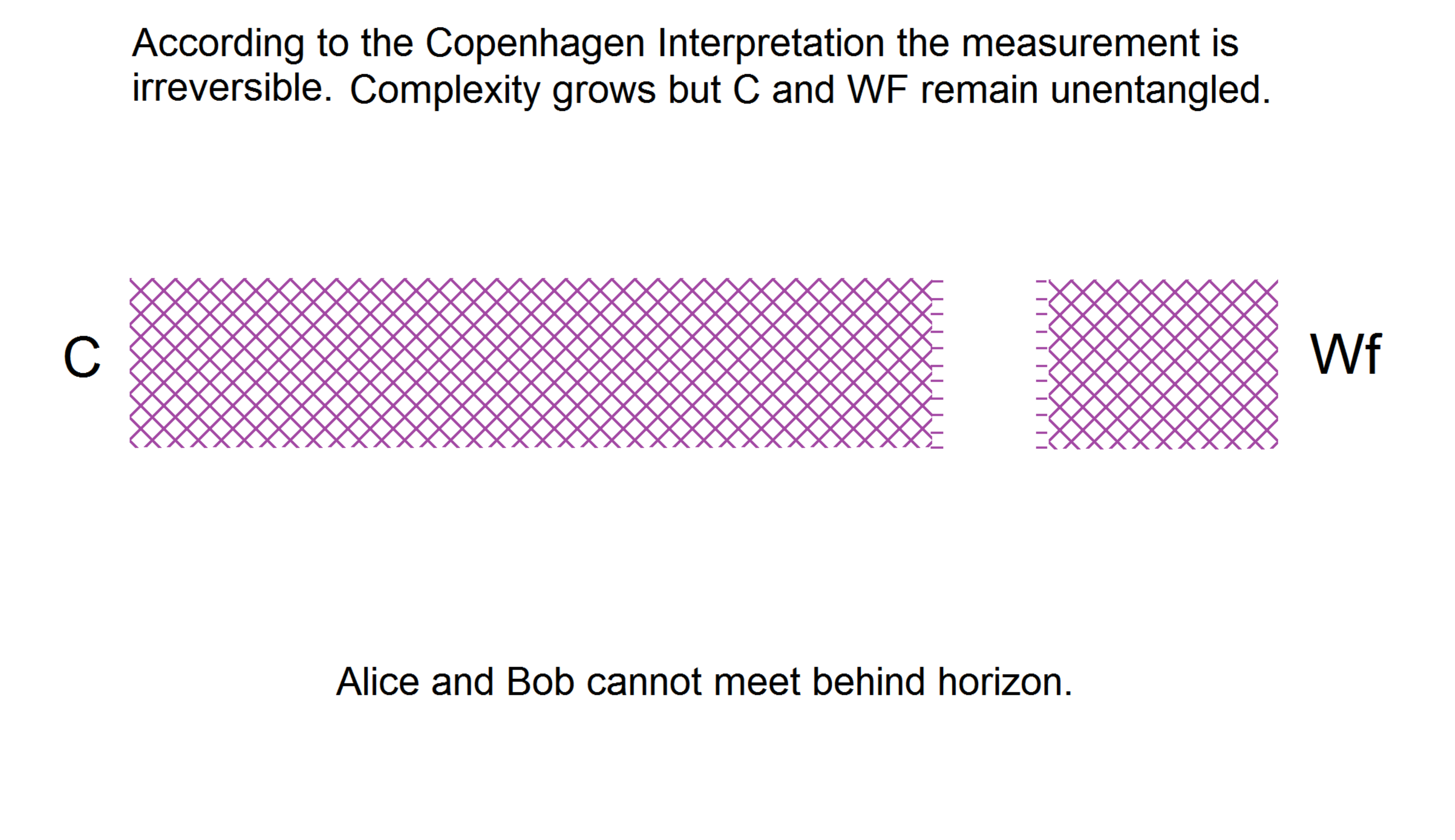}
\caption{}
\label{f13}
\end{center}
\end{figure}

Einstein, still being out of the room and not having made any observation, describes everything inside the room as a single quantum system---cat, Wigner's Friend, and Wigner. From the Relative State perspective 
no collapse occurred  when Wigner observed his friend; instead the three of them entered into some sort of tripartite entangled state in which   the cat would be entangled with the union of Wigner and his friend ( \ $W\cup WF$ \ ). Einstein concludes that when Wigner is  compressed to a black hole, there must be an ERB  connecting all three---the cat, Wigner, and Wigner's Friend. He also maintains that  messages from the cat-end can meet messages from the $W\cup WF$-end, inside the ERB.
\begin{figure}[H]
\begin{center}
\includegraphics[scale=.4]{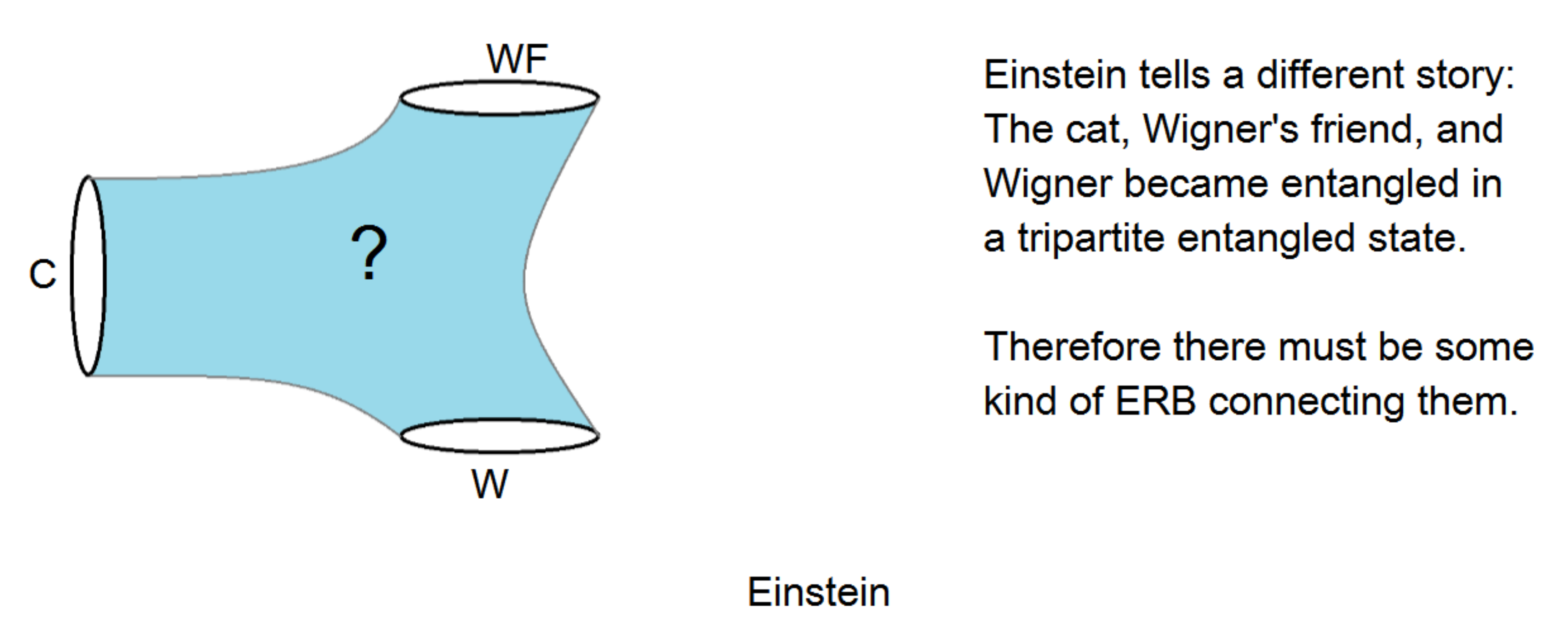}
\caption{}
\label{f14}
\end{center}
\end{figure}

It seems there is an apparent contradiction; the Copenhagen Interpretation says that no messages can pass between the cat, and either Wigner, or  Wigner's Friend.  
 But Einstein, using the \RSF insists that a  message can be sent from the cat to the union of Wigner and his friend. To resolve the contradiction \cite{Susskind:2014yaa} we need to know a little about tripartite entanglement.

\bn

\begin{figure}[H]
\begin{center}
\includegraphics[scale=.3]{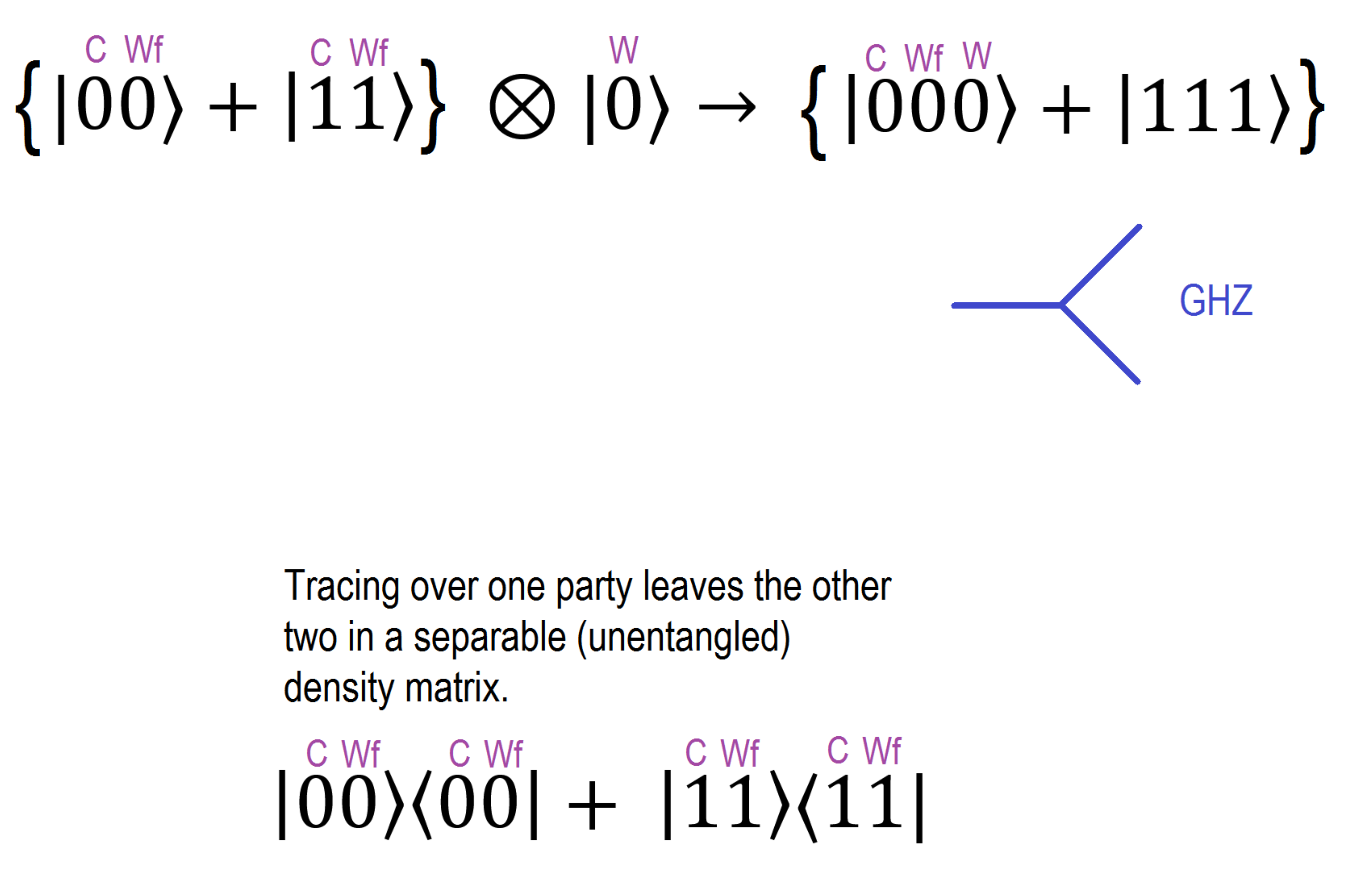}
\caption{}
\label{f16}
\end{center}
\end{figure}

Let's  simplify everything by replacing the cat, Wigner's Friend and Wigner by single qubits.  Go back to the point where Wigner's  Friend and the cat are entangled, and Wigner is about to do a measurement on his friend. Wigner begins in state $|0\ra.$ If he sees his friend in state $|0\ra$ he remains in state $|0\ra.$ If he sees his friend in state $|1\ra$ Wigner transitions to state $|1\ra.$ The result is a GHZ state $|000\ra+|111\ra.$
Figure \ref{f16} illustrates the process.
The small three-pronged figure  is a tensor network for the GHZ state. It has the property that it is non-zero only if all three external lines are the same---either $0$ or $1.$ In that case the value of the tensor is $1.$ We can express it in the form,
\be
T_{ijk}= \delta_{ij}\delta_{ik}=\delta_{ij}\delta_{ik}\delta_{jk}
\ \ \ \ \ \ \  \rm {no \ sum}
\ee

\bn
If one of the three parties in the GHZ state is traced over, the other two are left in a separable density matrix. That means that the density matrix is a sum of projection operators on unentangled pure states. In other words no two parties are entangled but any one party is maximally entangled with the union of the other two.

\begin{figure}[H]
\begin{center}
\includegraphics[scale=.3]{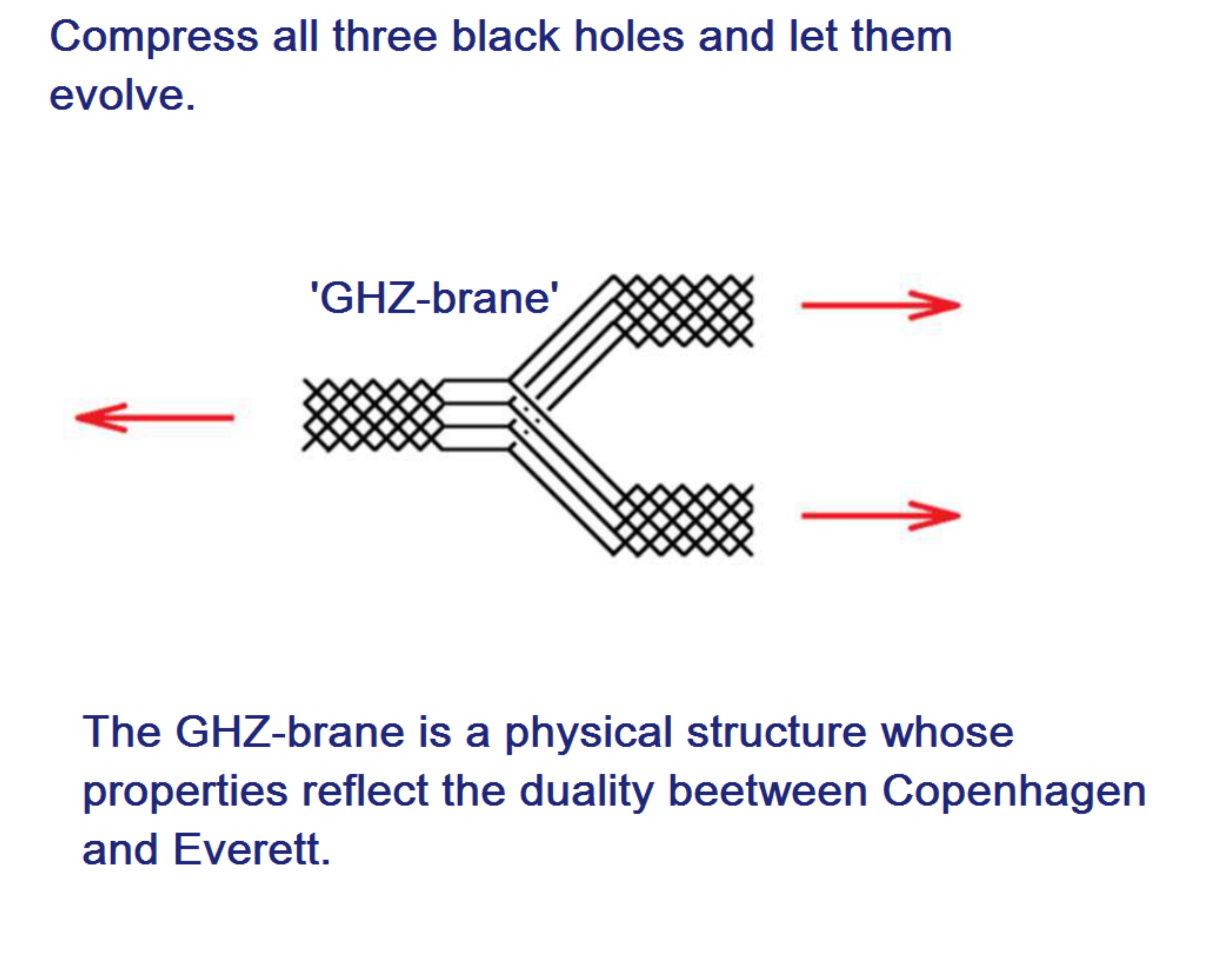}
\caption{}
\label{f18}
\end{center}
\end{figure}

We can generalize this to more complex systems modeled as collections of many qubits. Start with two such systems---the cat and Wigner's Friend---in a maximally entangled state. Meanwhile Wigner  (still outside the room) is in the state $|00000...\ra.$
Wigner enters the room and observes his friend in the Z basis. Each of the   qubits in Wigner's register becomes correlated with the friend's qubit, thus producing a product of GHZ triplets.
\begin{figure}[H]
\begin{center}
\includegraphics[scale=.5]{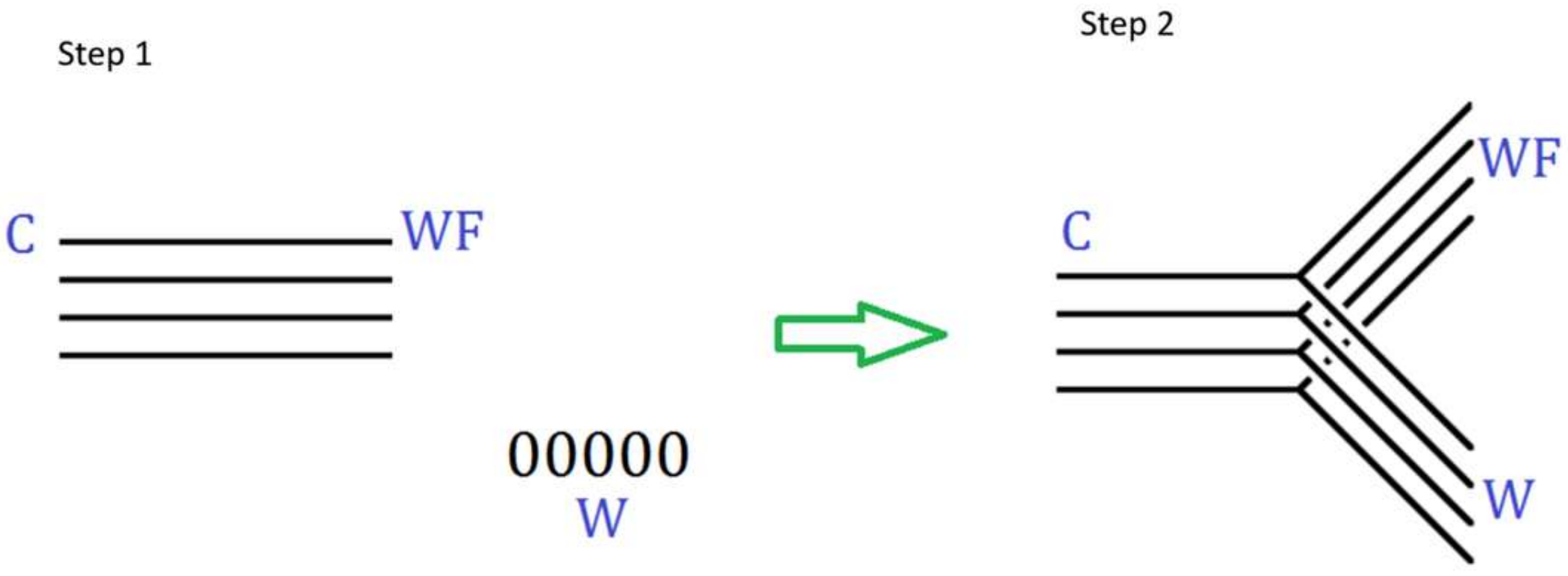}
\caption{}
\label{f17}
\end{center}
\end{figure}
By collapsing each of the three ends to form black holes and allowing them to evolve, some sort of ERB, linking the three parties, is created. That's shown as a tensor network in figure \ref{f18}.
Once the wormhole configuration has been created the black holes  will evolve and become increasingly complex \cite{Susskind:2014rva}\cite{Stanford:2014jda}. The three ends of the tripartite wormhole  grow linearly with time, but no  matter how large this tripartite wormhole grows the GHZ core at the center never disappears.  The existence of maximal GHZ entanglement is an invariant property of the ERB.

The  feature at the center of the ERB is a generalized geometric object that has not yet been studied. Its properties reflect the duality between the \CI and \RSF of quantum mechanics.  For want of a better name I've called it a GHZ-brane.

What are the properties of a GHZ-brane? First of all it is localized in the wormhole. Its properties  are not describable by classical geometry but it is a large localized object with a distillable GHZ entanglement of order the entanglement  entropy of the original black holes. It has the interesting property that it does not allow messages from any two parties to meet in the interior. But since any one party is entangled with the union of the other two, if Wigner and his friend cooperate, they can jointly send a message  that  can meet  another message launched by the cat. 
\begin{figure}[H]
\begin{center}
\includegraphics[scale=.3]{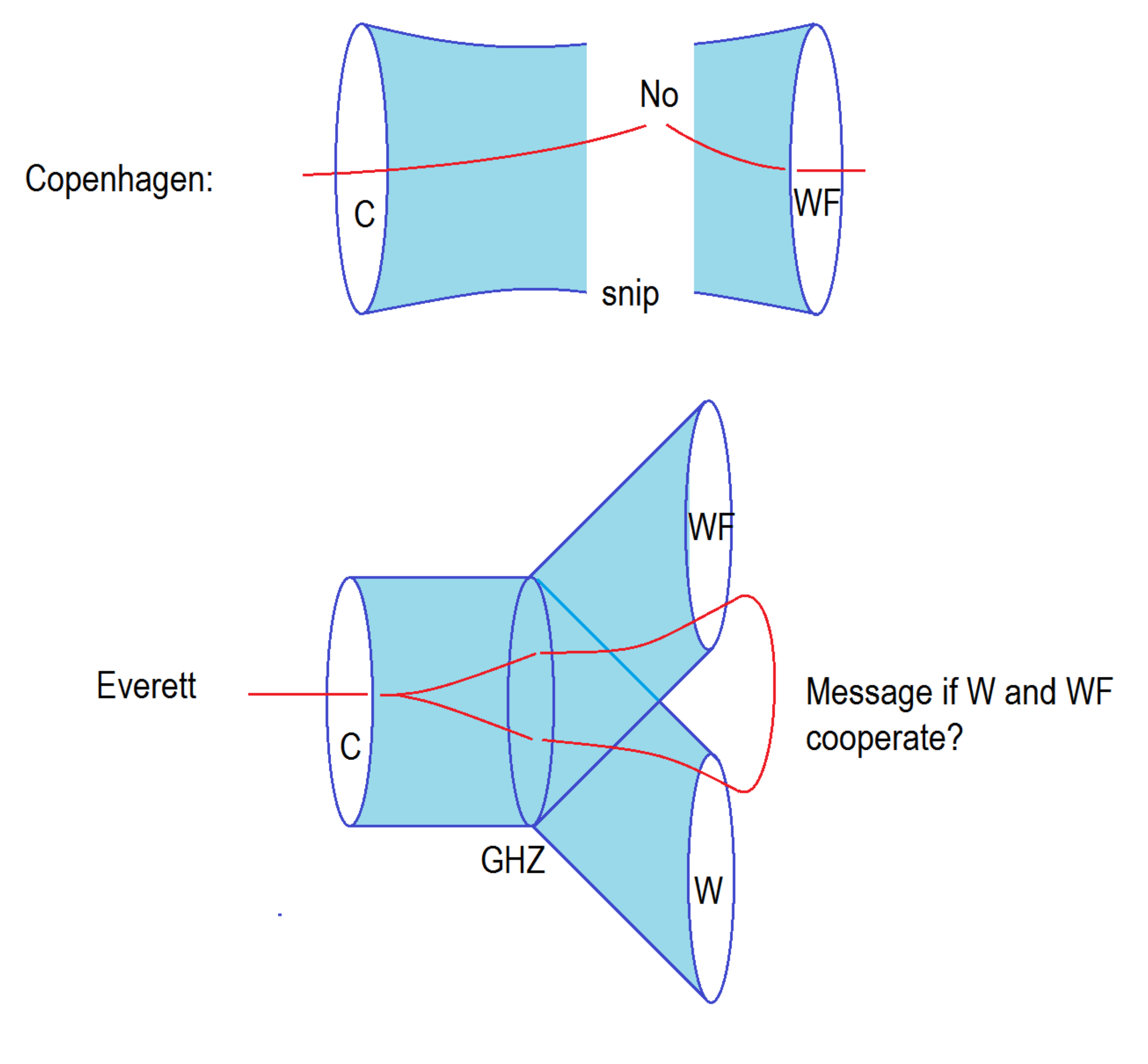}
\caption{}
\label{f21}
\end{center}
\end{figure}

Let's consider a protocol to do just that. Here Einstein will be helpful. Suppose that when Einstein enters the room he makes a measurement of Wigner's qubits. If he makes the measurement in the Z-basis he just confirms Wigner's observation. That's true in either the \CI or the Relative State Formulation. In the \RSF the final state of cat, friend, Wigner, and Einstein will be a four-party GHZ state, $|0000\ra+|1111\ra$. 

 But there are possibilities inherent in quantum mechanics that are lost in the \CI  when we say that the collapse of the wave function is irreversible. In order to bring these possibilities to light, Einstein does something unexpected: he measures Wigner's qubits in the X basis instead of the Z basis. 

\begin{figure}[H]
\begin{center}
\includegraphics[scale=.22]{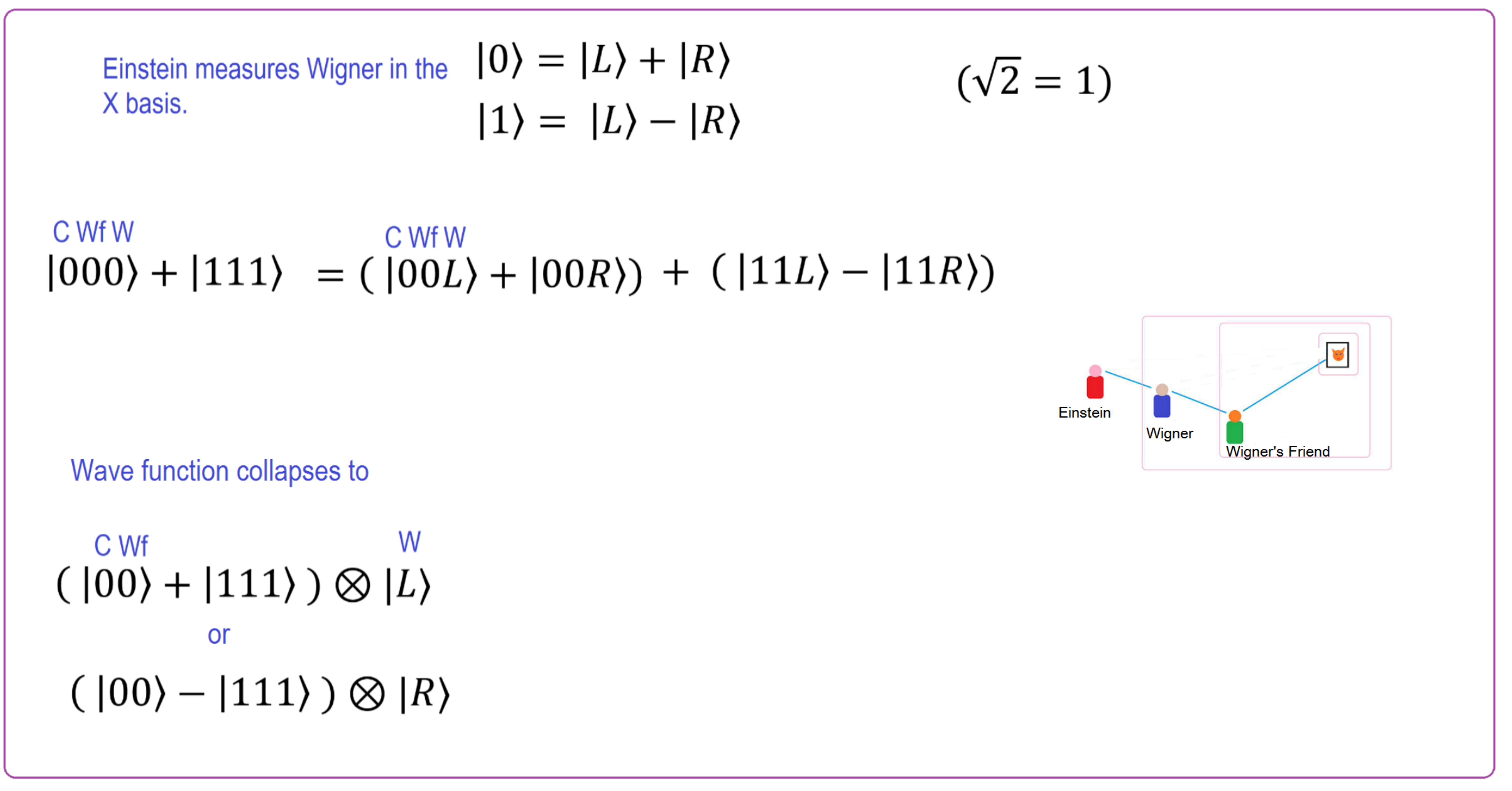}
\caption{}
\label{f22}
\end{center}
\end{figure}

To see what happens let's first  replace all parties by single qubits.  The states in the Z basis are called $|0\ra$ and $|1\ra.$ In the X basis I'll call them $|L\ra$  and  $|R\ra$ (for left and right). Ignoring factors of $\sqrt{2},$
\bea
|0\ra \eq |L\ra + |R\ra \cr \cr
|1\ra \eq |L\ra - |R\ra
\eea
Figure \ref{f22} outlines the analysis.  

\bn

\bn

Einstein's measurement can yield one of two  results: $X=L$ or $X=R.$ The interesting thing is that in either case the state of the  cat/friend system is projected back to a maximally entangled state. In fact if $X=L$ the cat/friend state is projected back to the original entangled state $|00\ra +|11\ra$ (the state before Wigner's measurement) and Einstein's job is done.

On the other hand if the result is $X=R$ the cat/friend state is projected to $|00\ra -|11\ra.$ This state  is maximally entangled, but it is not the original state. However there is a simple protocol that Einstein can apply to insure that the cat/friend state returns to $|00\ra +|11\ra.$ 
\begin{figure}[H]
\begin{center}
\includegraphics[scale=.3]{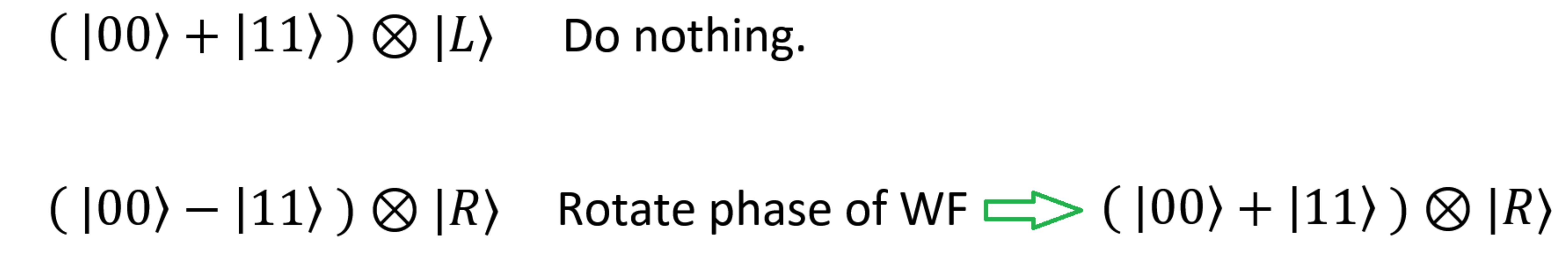}
\caption{}
\label{f23}
\end{center}
\end{figure}

\bn
If the outcome was $X=L$ Einstein does nothing. But if $X=R$ he executes a simple unitary operation on Wigner's Friend by acting with 
 \be
Z=  \left( \begin{smallmatrix} 1&0\\ 0&-1 \end{smallmatrix} \right).
\ee
This rotates the state
 $|00\ra -|11\ra$ to  the state $|00\ra +|11\ra.$ Thus whether the outcome of Einstein's measurement is $L$ or $R$ the cat/friend system ends up in the original maximally entangled state.

Note that the entire protocol involved actions only  on Wigner and Wigner's Friend. The cat was not involved at all. This shows that a  cooperative experiment at the Wigner/Wigner's-Friend side of the tripartite system can restore the original entangled state of the cat and Wigner's Friend.

\begin{figure}[H]
\begin{center}
\includegraphics[scale=.3]{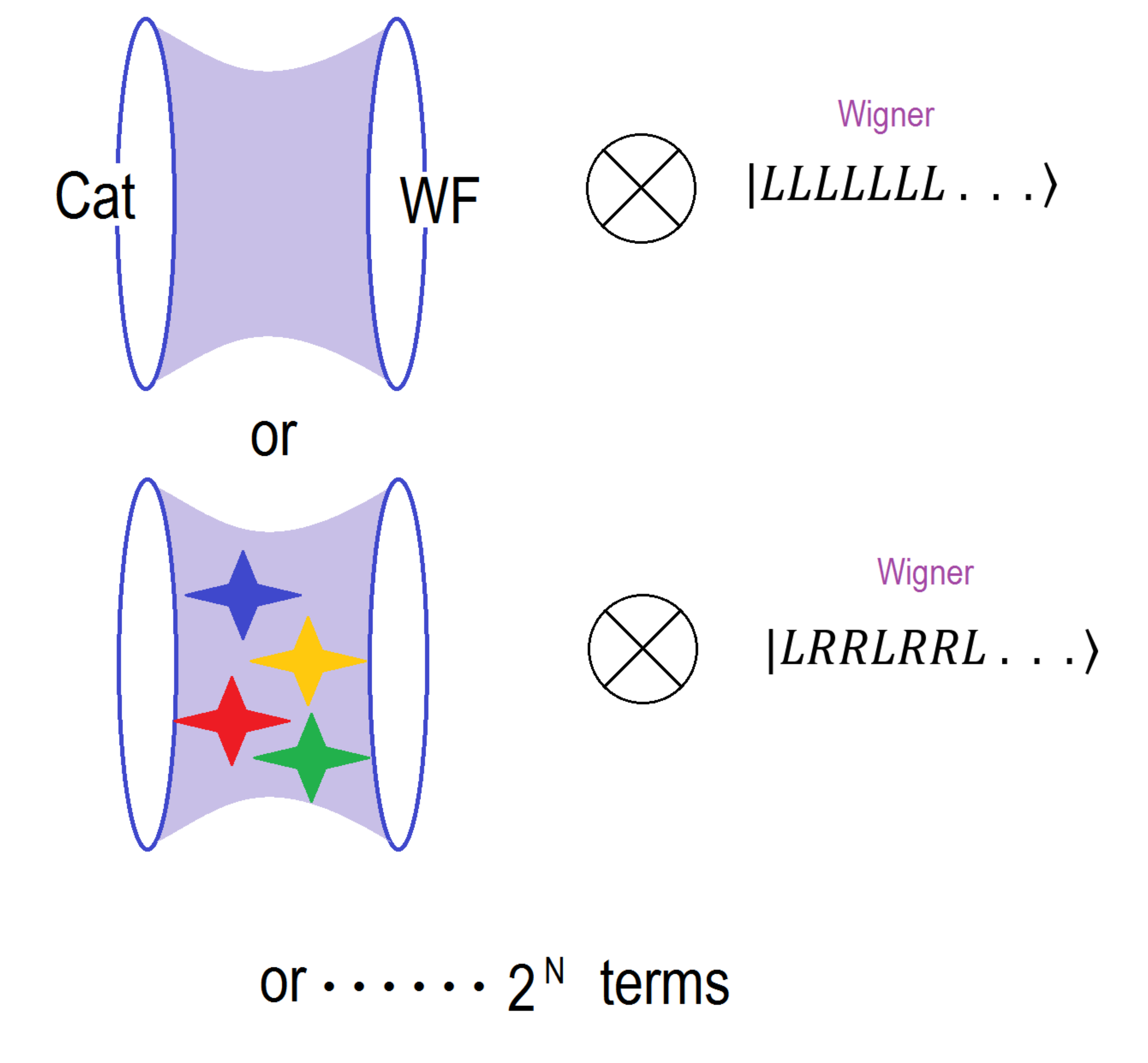}
\caption{}
\label{f24}
\end{center}
\end{figure}

\begin{figure}[H]
\begin{center}
\includegraphics[scale=.2]{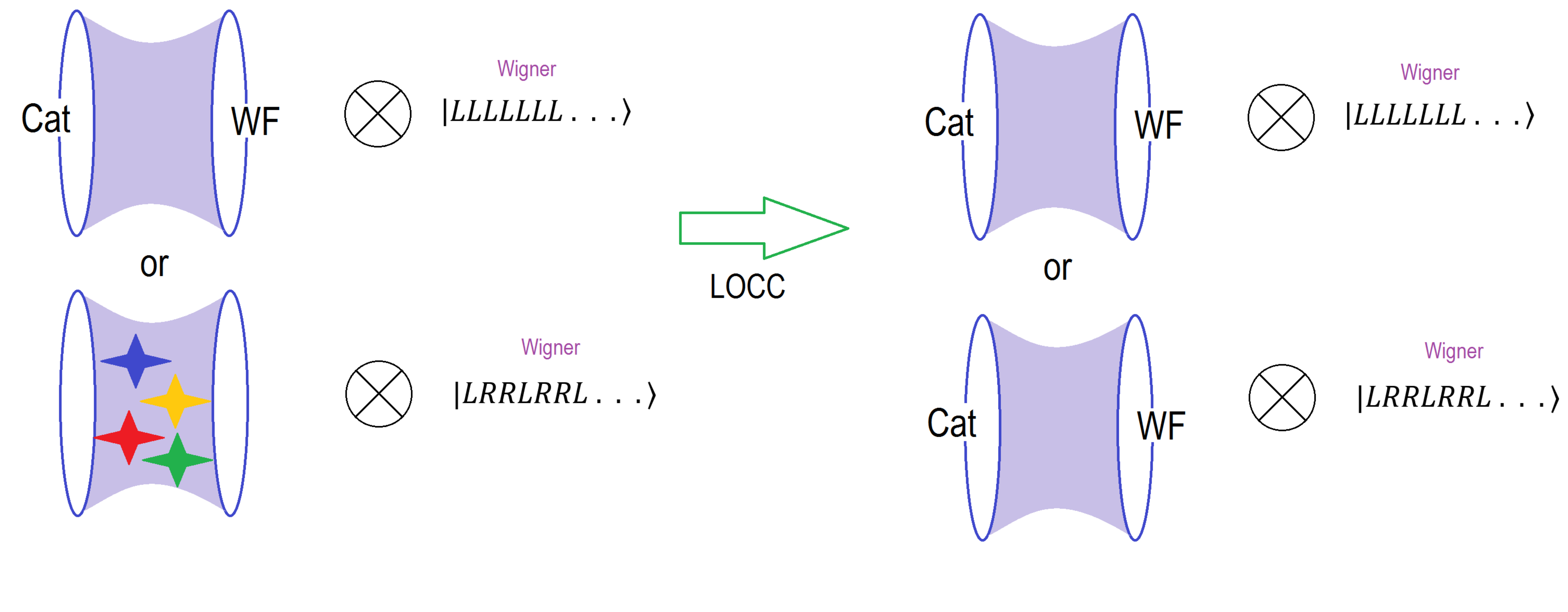}
\caption{}
\label{f25}
\end{center}
\end{figure}

The generalization to many qubits is straightforward and shown in figure \ref{f24} and \ref{f25}. 

\bn

\bn

\bn

\bn
Once the original entangled state is restored it is possible to send messages from the cat and Wigner's Friend that meet in the ERB.

\bn

\bf    
EXERCISE: You are a fourth observer outside the room.

\bn

a) What is your RSF description of the system  consisting of Einstein, Wigner, Wigner's Friend, and the Cat?  
\bn 

b) If Einstein is also compressed to a black hole, what is the description of Einstein, Wigner, Wigner's Friend and the cat, in terms of ERB's?

\rm

\bn 

The GHZ brane is a new object that could be said to reflect the duality between the \CI and the Relative State Formulation. 
It is a localized object in the spacetime behind the horizons of the three entangled black holes. It can  be constructed by a simple known procedure. Starting with a collection of a large number of GHZ-entangled triplets shared between Alice, Bob, and Charlie, the three shares are compressed to form a triplet of black holes. After some period of evolution the interior geometry will consist of three tube-like regions bound together by a GHZ brane at the center.

The amount of GHZ entanglement is invariant under local unitary transformations and is maximal for the GHZ brane. It would  be nice  to know how much GHZ entanglement is present in smooth classical tripartite wormholes \cite{Gharibyan:2013aha}\cite{Balasubramanian:2014hda}   like the one shown in figure 
\ref{f14}. This question is presently being studied by 
G. Salton, B. Swingle,  and M. Walter\footnote{To be published.}. The tentative conclusion is that 
 classical multipartite wormholes have very little or no GHZ entanglement. (A similar conclusion is suggested by the results of \cite{Gharibyan:2013aha}\cite{Balasubramanian:2014hda}.) If correct this would  show that GHZ-branes and tripartite wormholes with classical geometry are distinct objects that are distinguishable in an invariant way.

\begin{figure}[H]
\begin{center}
\includegraphics[scale=.3]{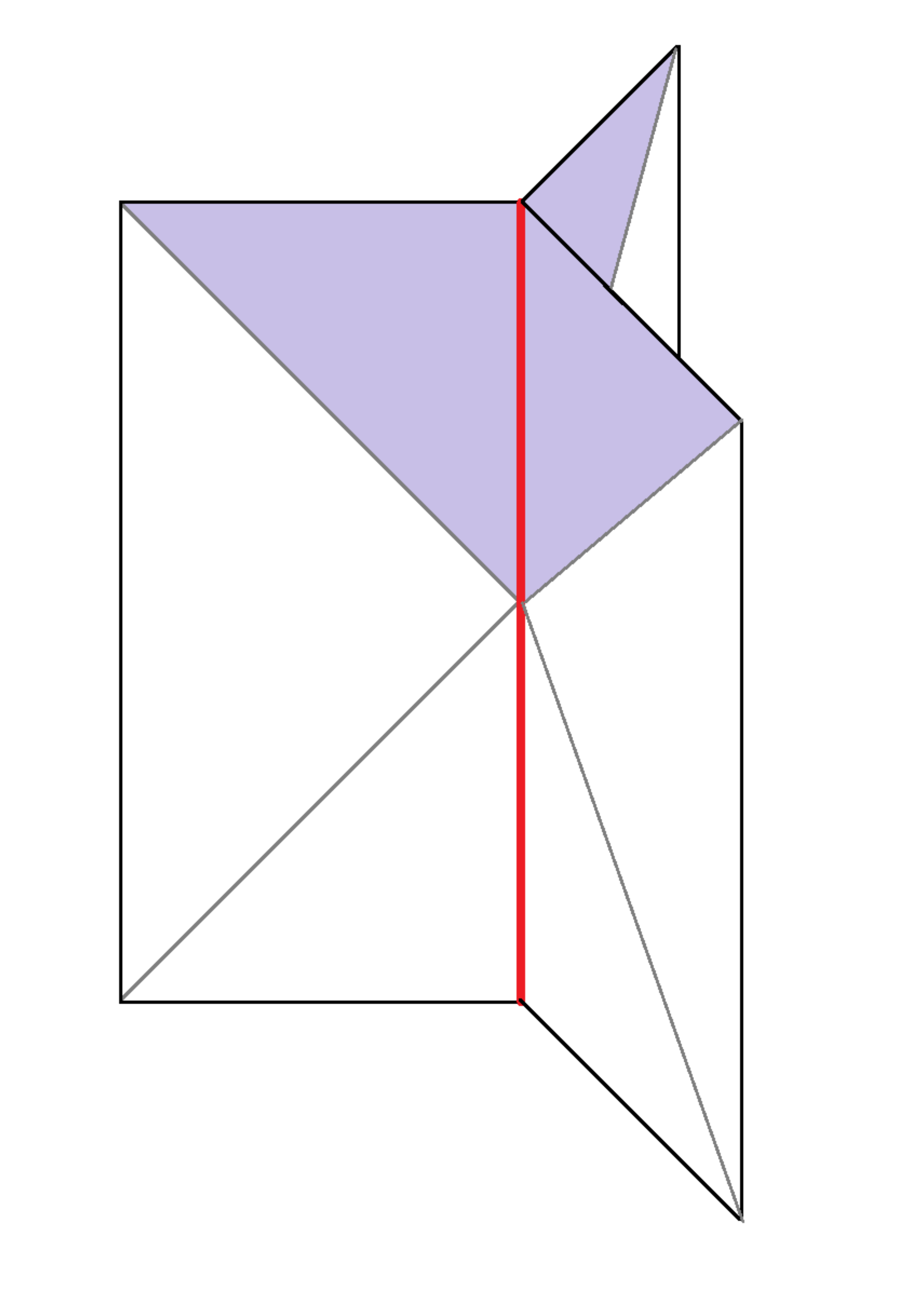}
\caption{}
\label{g4}
\end{center}
\end{figure}

The overall spacetime geometry of a GHZ-brane would have the structure shown in figure \ref{g4}.
\bn


\sc
\section{Teleportation Through the Wormhole: ERBs as a Resource}

Earlier I said that ERBs are part of a resource, but I haven't explained how the resource can be used. In this second part of the lecture I will show you how complex quantum information can be teleported through an ERB \cite{Susskind:2014yaa}.  Similar ideas appeared in a very interesting recent paper by    Numasawa, Shiba, Takayanagi and Watanabe\cite{Numasawa:2016emc}  \ as well as in earlier work by Marolf and Wall \cite{Marolf:2012xe}.

The goal of teleportation is not to transmit information faster than the speed of light; that's of course impossible. The goal is to transfer information in a way that does not allow anyone to intercept it and reconstruct what was teleported.

We begin with Alice and Bob each in control of two unentangled but similar  black holes, $A$ and $B.$ We assume Alice and Bob are very far from one another.

\begin{figure}[H]
\begin{center}
\includegraphics[scale=.5]{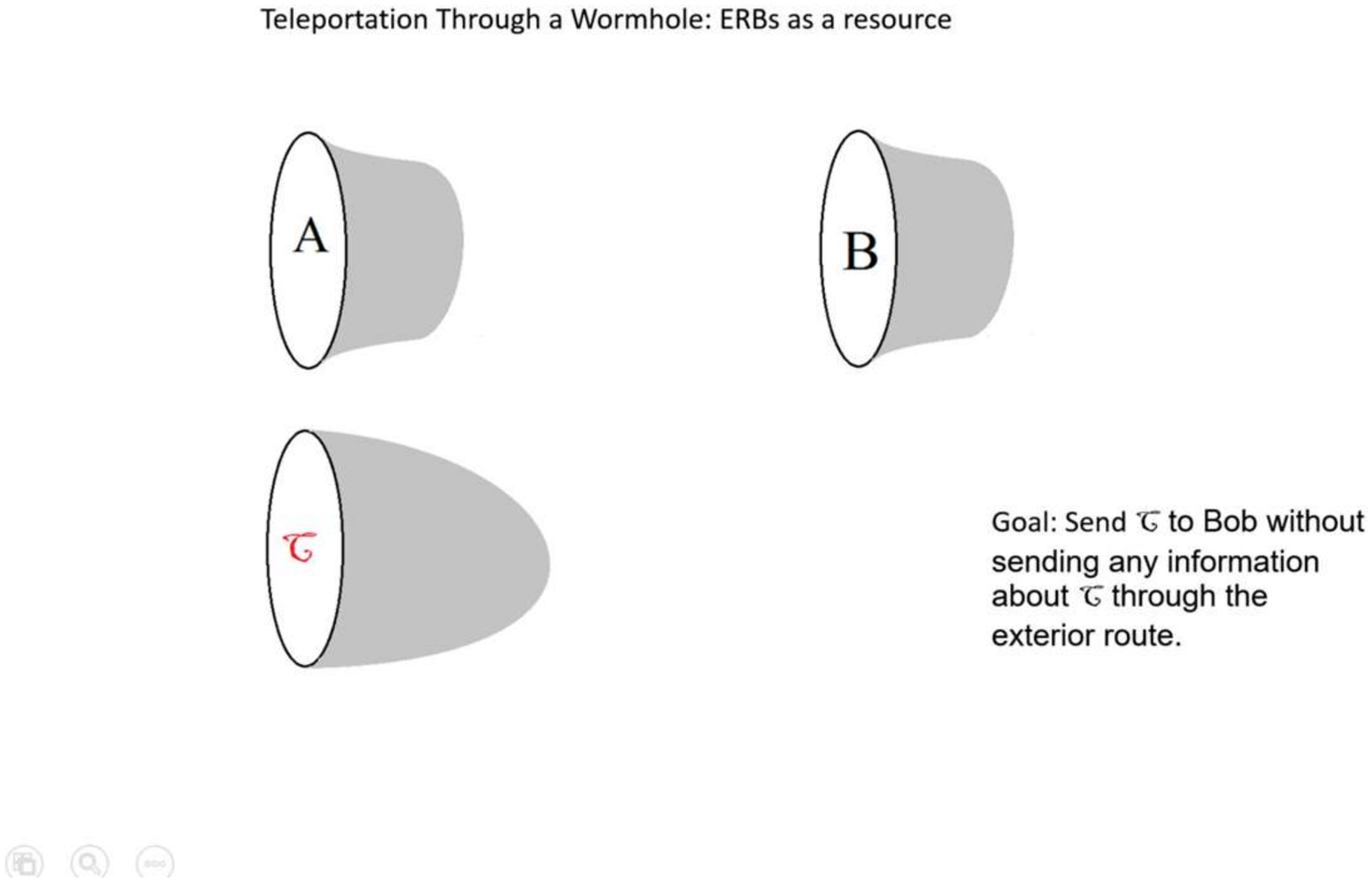}
\caption{}
\label{f28}
\end{center}
\end{figure}

In addition there is a third system shown as a fancy red $\cal{C}$ in figure \ref{f28}. 
The figure depicts  $\cal{C}$ as another black hole of the same size as $A$ and $B,$ but it could be anything as long as it is not bigger in information content than $A$ and $B.$ Alice's goal is to send the quantum state $\cal{C}$ so that it cannot be intercepted.  She can of course just send $\cal{C}$ to $B,$ but in that case if the message is intercepted, the thief can learn  about $\cal{C}.$ 

In fact there is no way that Alice can accomplish her goal unless she and Bob share a sufficiently large resource of entanglement. Let's suppose that they do share such a resource in the form of an ERB connecting the black holes.
\begin{figure}[H]
\begin{center}
\includegraphics[scale=.5]{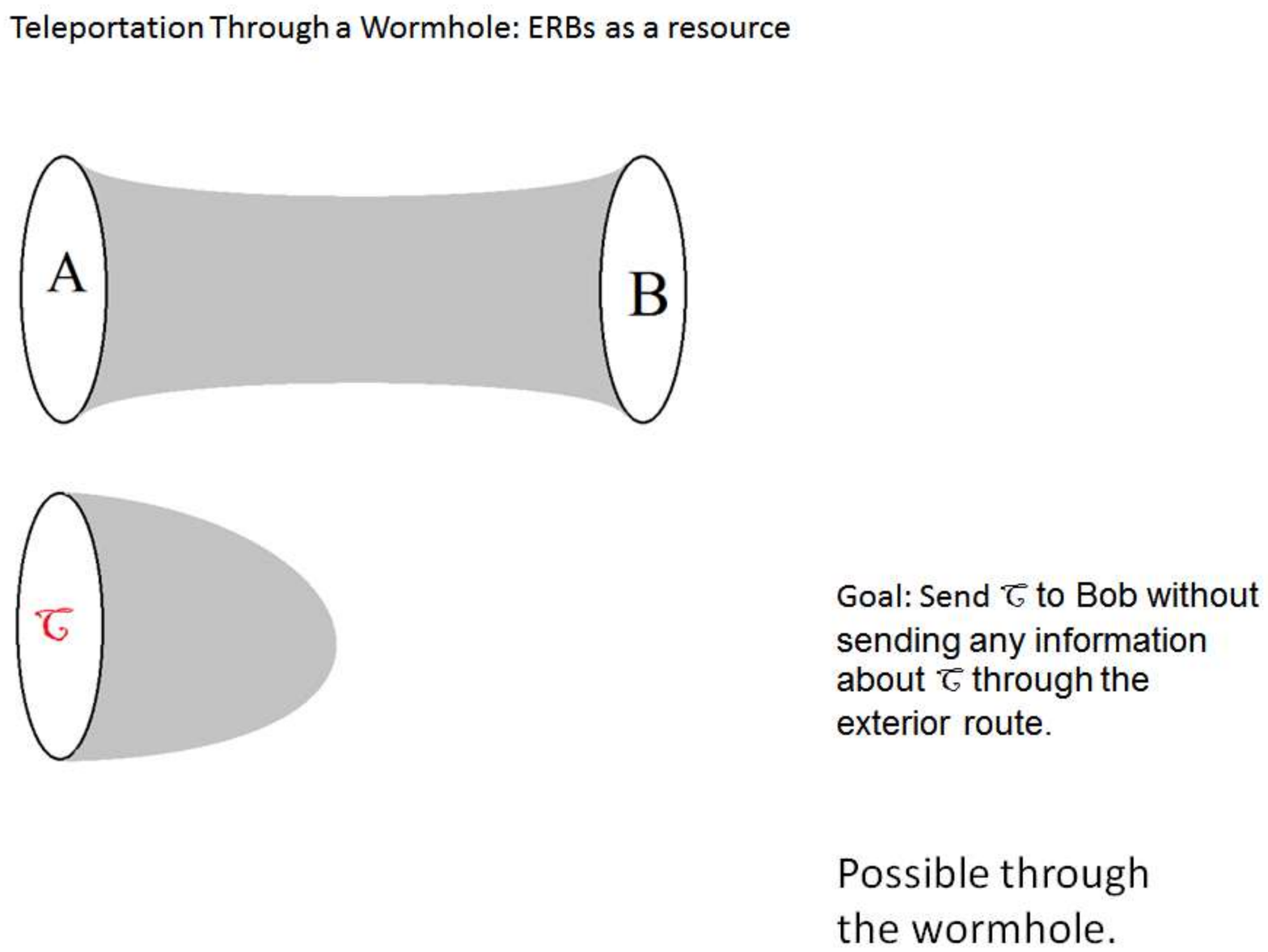}
\caption{}
\label{f29}
\end{center}
\end{figure}
\bn
I will describe the protocol without proving that it works. You can find the analysis in \cite{Susskind:2014yaa}.
The first step is to throw $\cal{C}$ into $A$ to form a black hole of twice the information in $\cal{C}.$ 
\begin{figure}[H]
\begin{center}
\includegraphics[scale=.5]{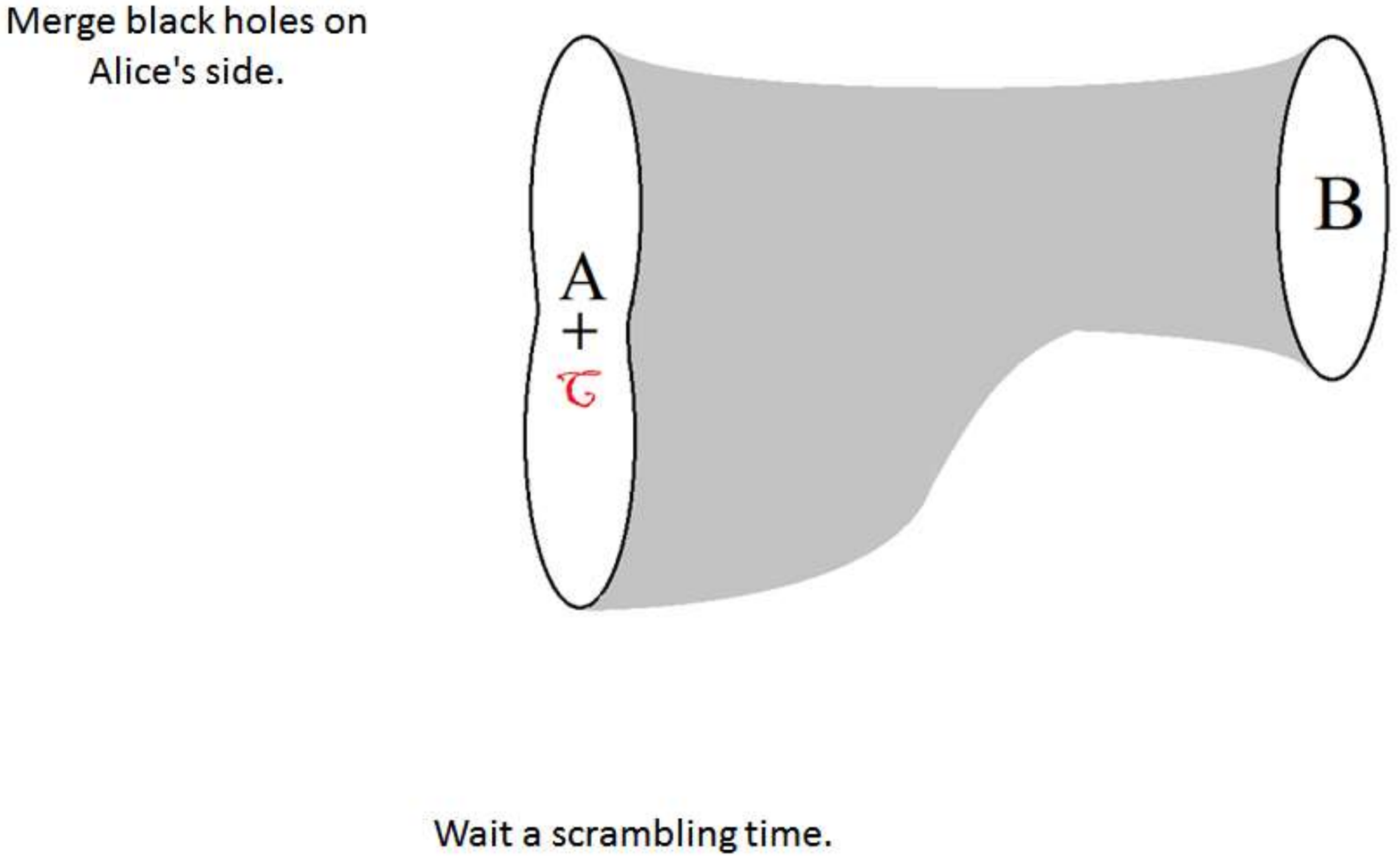}
\caption{}
\label{f30}
\end{center}
\end{figure}

\bn
Then  wait a scrambling time \cite{Hayden:2007cs}\cite{Sekino:2008he} in order that the information in $\cal{C}$ and $A$ gets thoroughly mixed.
Once the information in $\cal{C}$ and $A$ has been scrambled Alice makes a complete measurement---say in the Z basis---of all the qubits comprising her new black hole. This snips off her black hole and leaves ${\cal{C}}'s$ information trapped behind Bob's horizon. 
\begin{figure}[H]
\begin{center}
\includegraphics[scale=.5]{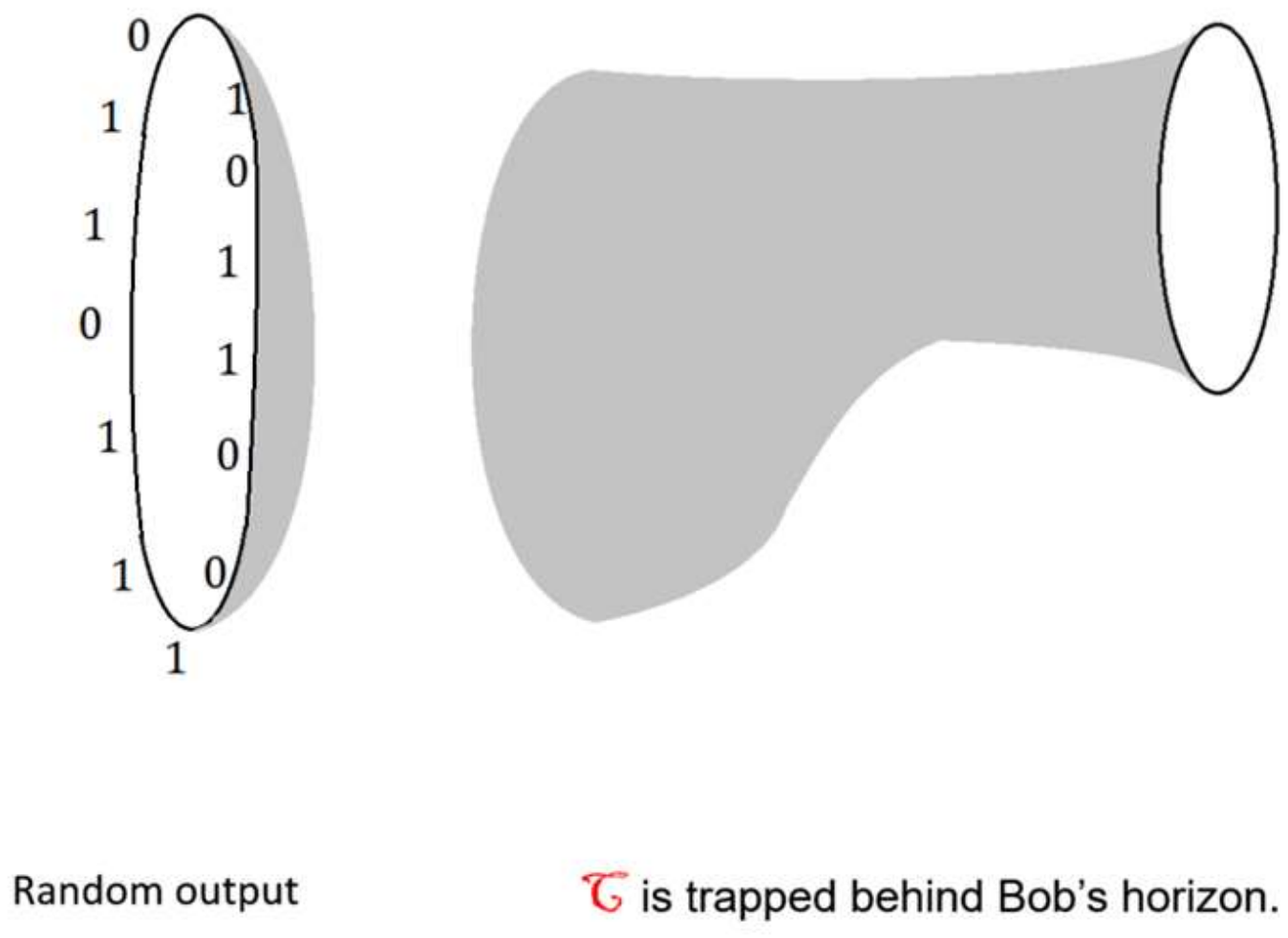}
\caption{}
\label{f31}
\end{center}
\end{figure}

The output of Alice's measurement is a collection of classical bits that she can record in her notebook. The important point is that the classical message in Alice's notebook is completely uncorrelated with the information originally carried by $\cal{C}.$ That independence is a result of the scrambling.

Next Alice sends  her notebook to Bob. 
\begin{figure}[H]
\begin{center}
\includegraphics[scale=.3]{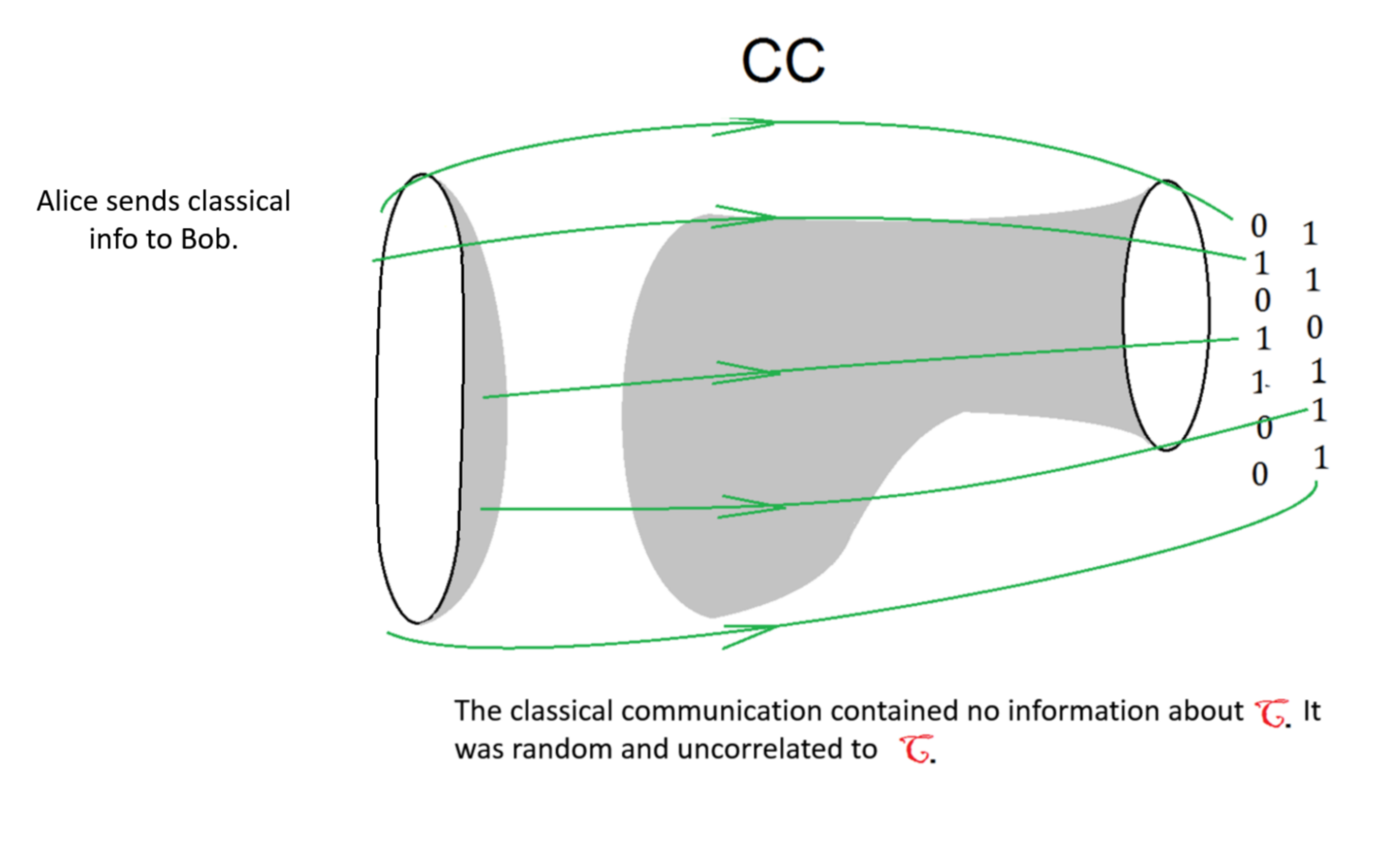}
\caption{}
\label{f32}
\end{center}
\end{figure}
\bn
This of course takes time; we are not trying to send $\cal{C}$ faster than light, only to get $\cal{C}$ to Bob with no possibility of interception.

Now comes the hard part of the protocol. When Bob opens the notebook he executes an operation on his black hole that depends on what he reads. If the notebook says $00000000000...$
he does nothing. If the notebook reads $0011101001010...$ he applies a specific unitary rotation to the qubits comprising his black hole. If Bob carries this out correctly then  he will find that the state of his black hole has been transformed into the original state of $\cal{C}.$

\begin{figure}[H]
\begin{center}
\includegraphics[scale=.5]{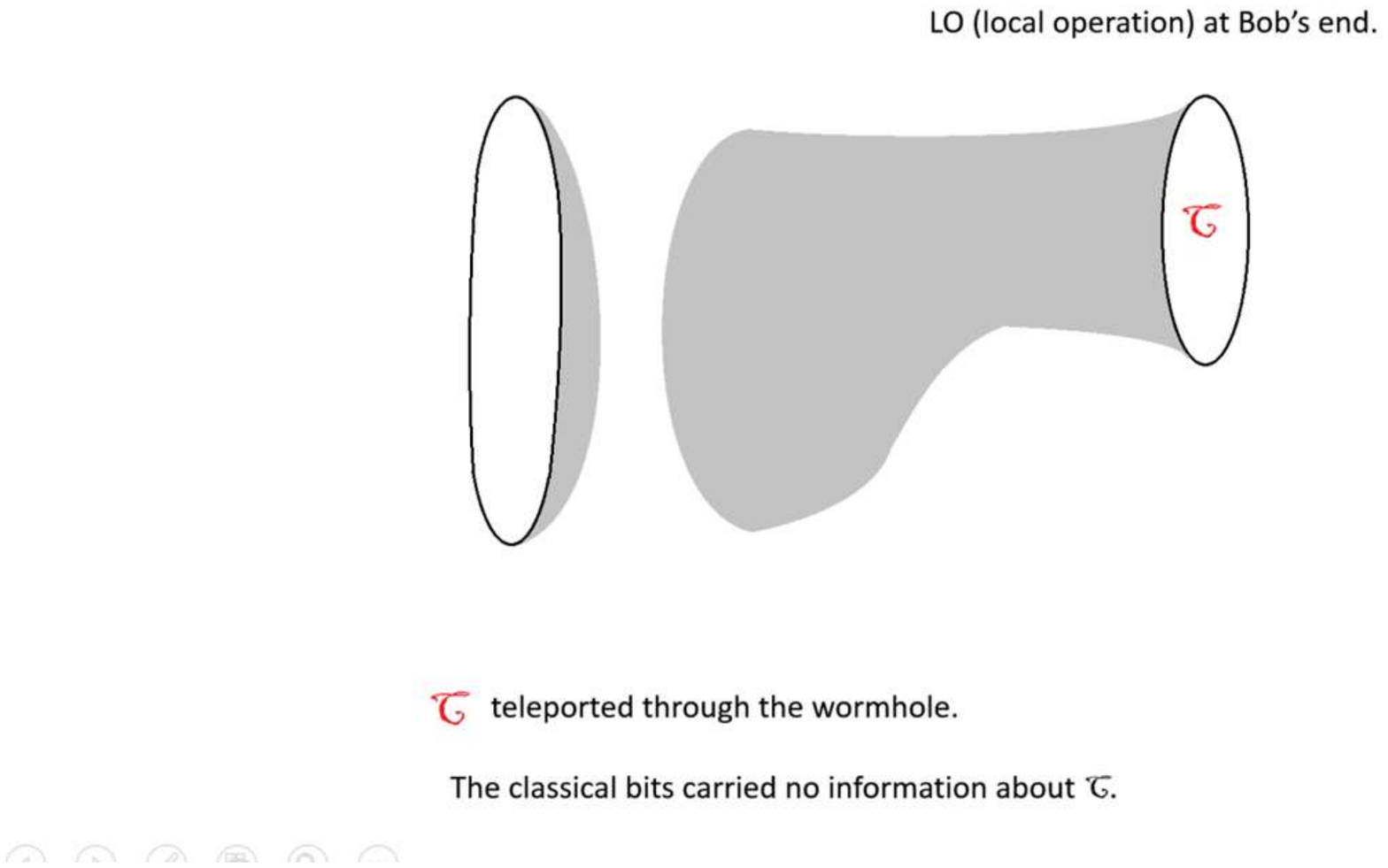}
\caption{}
\label{f33}
\end{center}
\end{figure}

Here is another picture illustrating the teleportation of 
$\cal{C}.$ 
\begin{figure}[H]
\begin{center}
\includegraphics[scale=.5]{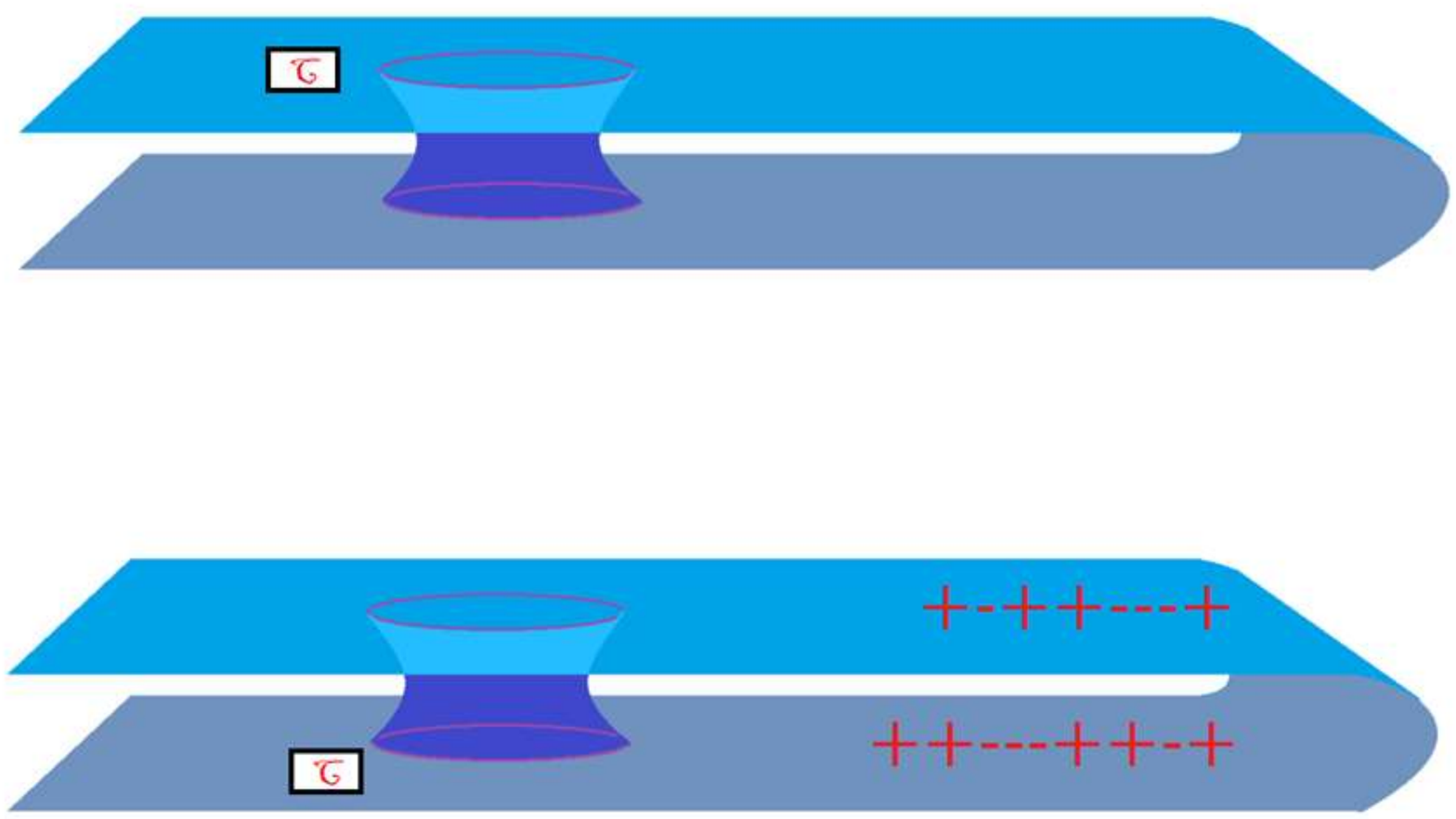}
\caption{}
\label{f34}
\end{center}
\end{figure}
The surprising  thing is not that $\cal{C}$ was sent instantaneously  (it was not), but that the classical bit-string that went the  ``long way" carried no information at all about $\cal{C}$. That information went through the wormhole.

One could say that there really is nothing new here; it's just conventional quantum teleportation  \cite{Bennett}\cite{Wooters} . Indeed it is,  but that misses the point. Suppose that after successfully teleporting $\cal{C}$ Alice and Bob conclude that there must be a wormhole connecting their black holes. Bob sends a classical message to Alice telling her that he will jump into his black hole if she will jump in to hers.
 With suitable preparation they will meet and indeed discover that the black holes are connected by an Einstein-Rosen bridge. There is a correlation: no bridge---no teleportation.

\sc
\section{Two Slits and a Wormhole}
Some would say that the essence of quantum mechanics is entanglement. Others would say that it is the interference of probabity amplitudes, an example being the two-slit experiment. 
Let me give a straw man argument that they are completely different things---interference and entanglement. Interference is a single particle phenomenon. It can be exhibited by independent particles which are sent through the apparatus one at a time. The time interval between them can be arbitrarily large and they can come from different sources which have never interacted. Entanglement by contrast is a multiparticle phenomenon that concerns particles that originate from a common source.
As a consequence, one would not expect interference phenomena to have anything to do with ER=EPR. 

\begin{figure}[H]
\begin{center}
\includegraphics[scale=.3]{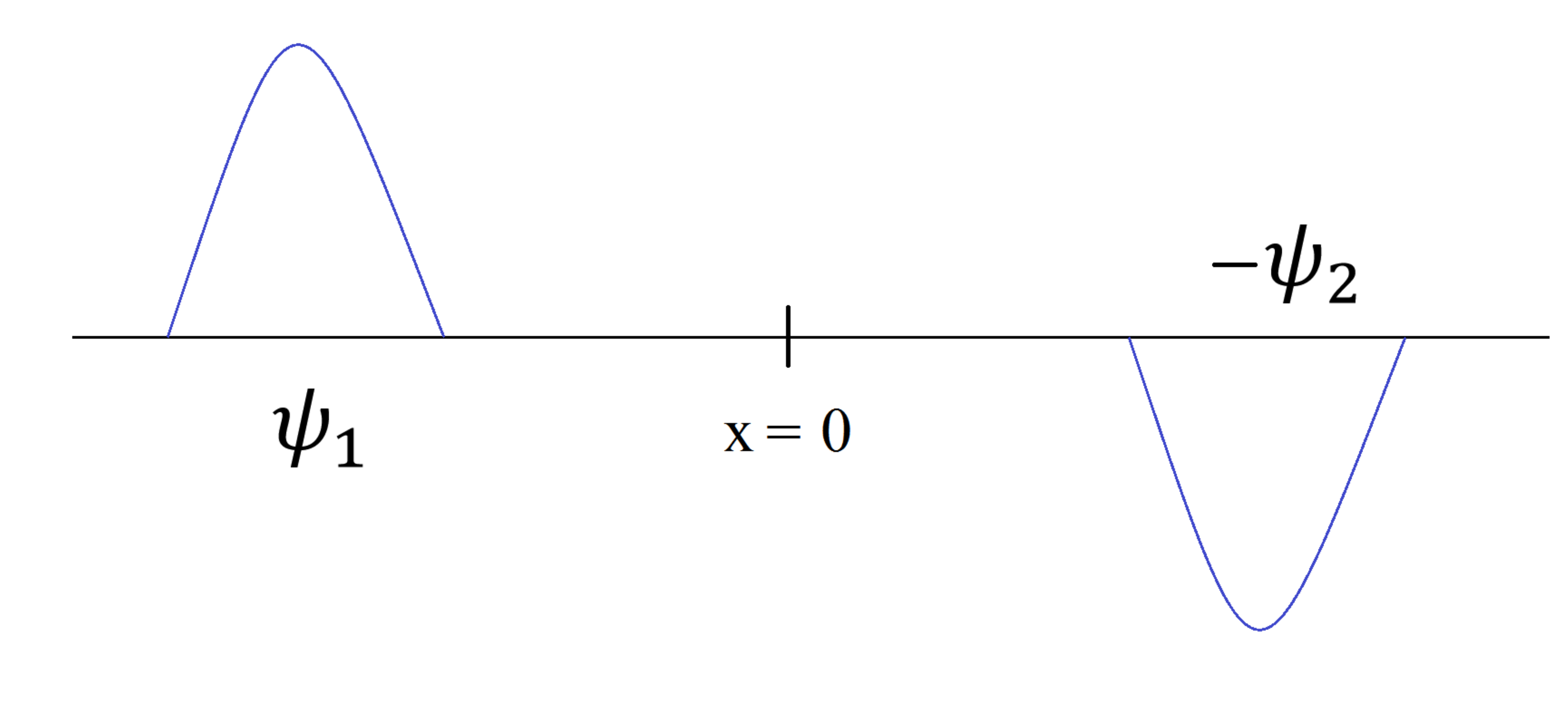}
\caption{}
\label{g2}
\end{center}
\end{figure}

This is incorrect and to see why
consider a single particle (to be definite let it be a fermion) in a superposition of states composed of  two non-overlapping wave packets with a relative phase. The packets could result from the particle passing through two slits or a half-silvered mirror.
For example in figure \ref{g2} I've drawn a wave function consisting of two non-overlapping packets with opposite sign. We can assume they are approaching each other and will soon overlap.
\bn
 If we allow the wave function to evolve there will be nodes at points of destructive interference.   Either wave packet by itself would allow the particle to appear at those points, but  allowing both ways of getting there would forbid a particle from showing up at a node. From a classical particle perspective this is strange. In the two-slit experiment  it leads to questions like, ``how did the particle going through one slit know that the other slit was open?"  The answer we learn in a course on quantum mechanics is quantum interference.

\begin{figure}[H]
\begin{center}
\includegraphics[scale=.3]{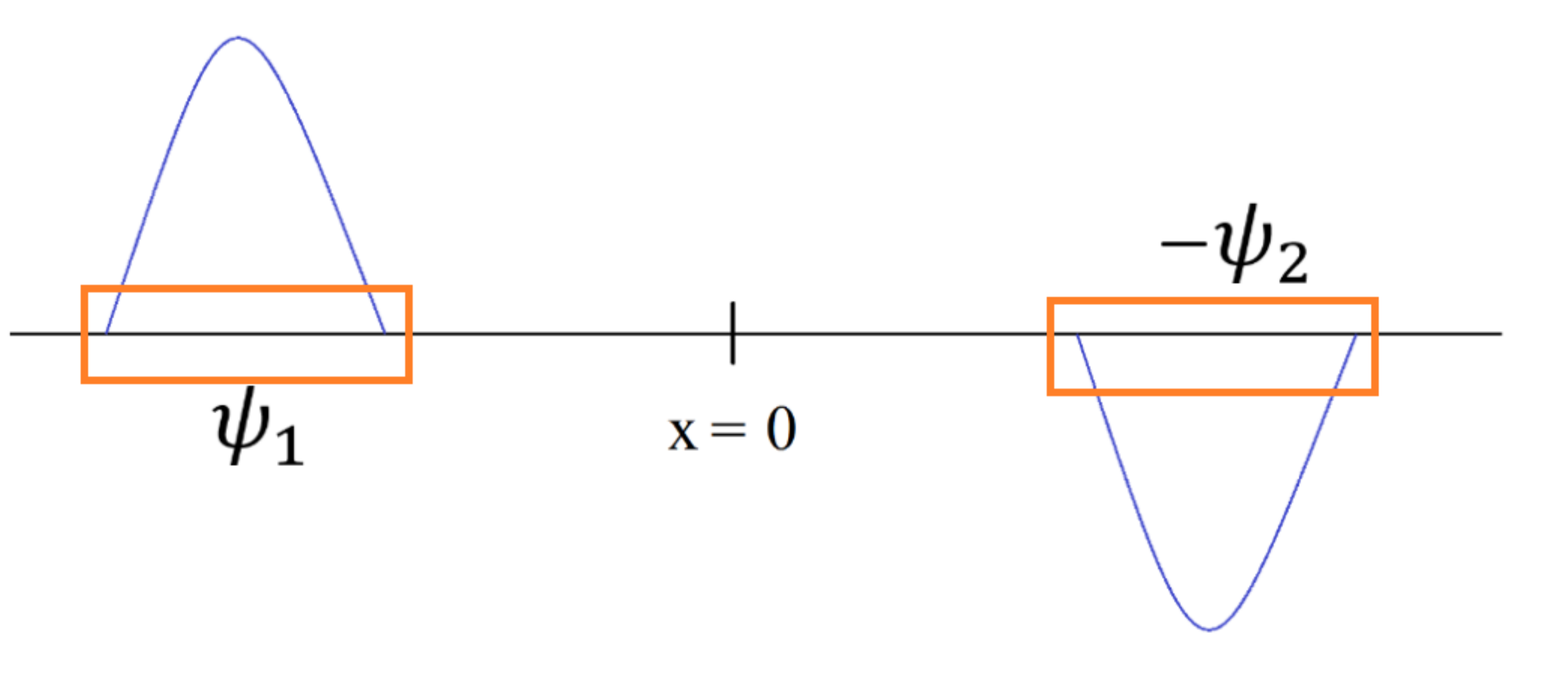}
\caption{}
\label{g5}
\end{center}
\end{figure}

Instead of focusing on the particle in the above example let's use  second-quantization and focus on the degrees of freedom in the two spatial boxes, shown in figure \ref{g5}. The boxes each  contain one of  the  packets.
The state represented by the wave function $\psi_1 - \psi_2$  \ may be re-expressed as
\be
|10\ra - |01\rangle
\label{one-ent}
\ee
where $|10\ra$ represents one particle in the left box, no particle in the right box; and $|01\ra$ represents no particle in the left box, one particle in the right box. It is evident from the form of \ref{one-ent} that the quantum fields in the two boxes are maximally entangled qubits. Focusing on the boxes, the state has the form of a Bell pair.

Now let's speculate that there exists an  ambitious version of ER=EPR. Assume that ER=EPR holds right down to the level of a single Bell pair. If that can be made precise  the two boxes in figure \ref{g5} would be connected by a Planckian ERB. 
Moreover as the  two wave packets approach each other the ERB would follow the wave packets ``reminding" each wave packet that the other is there.
The state of the ERB would reflect the relative phase; for example whether the packets were added with a plus or minus sign.

Evidently the nonlocal feature of quantum mechanics that we ordinarily call interference can be a special case of the nonlocality of entanglement. If we believe in the ambitious form of ER=EPR, this implies the presence of an Einstein-Rosen bridge connecting the superposed wave packets for a single particle.  The properties of the ERB reflect the type of interference, i.e., destructive or constructive.

Is there an experiment that can confirm the existence of ERBs in interference experiments? For that purpose
 consider the state of a particle just after it has passed through a 2-slit apparatus. It will be in a superposition of two packets representing the paths through the two slits.  
\begin{figure}[H]
\begin{center}
\includegraphics[scale=.3]{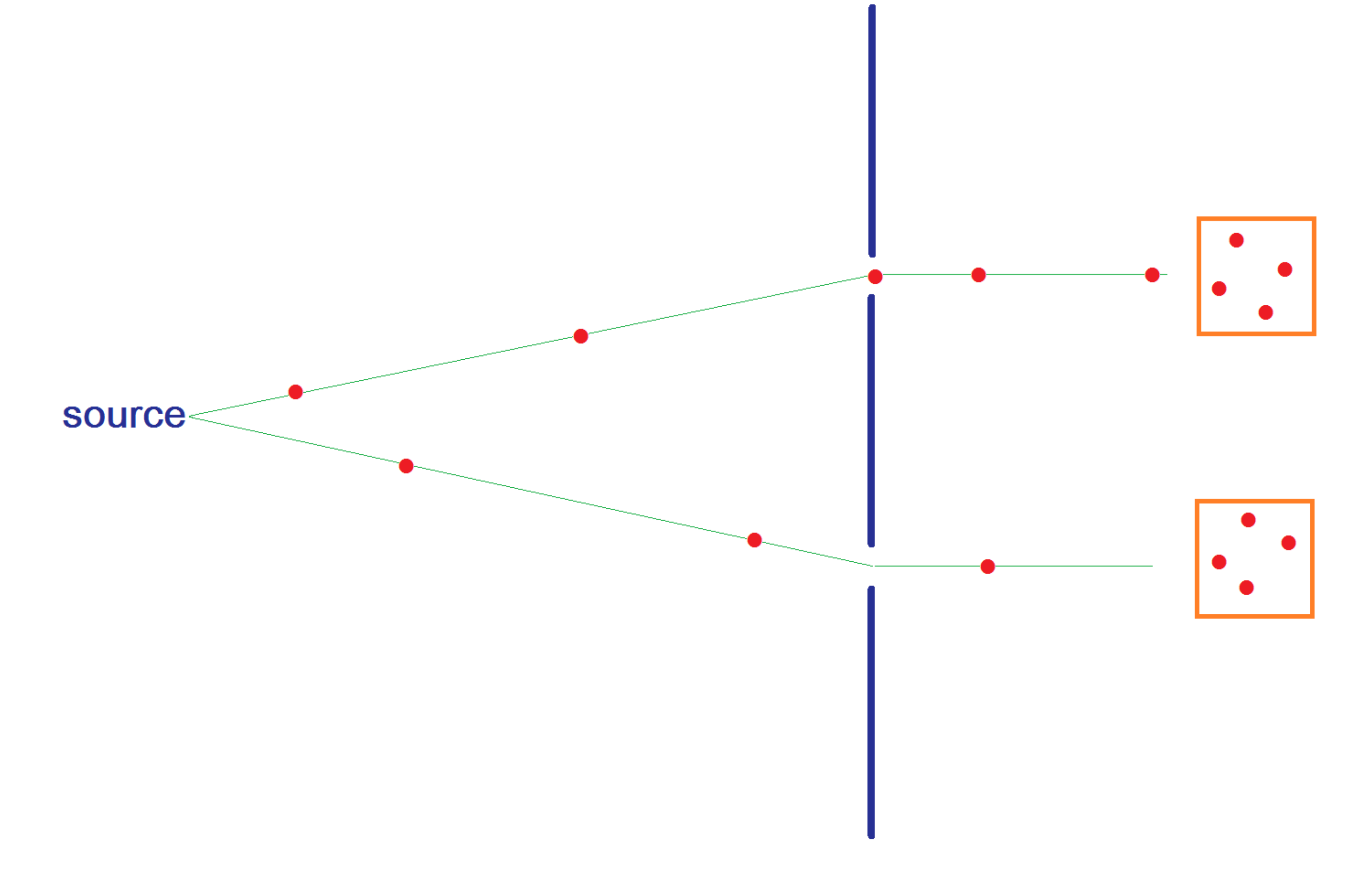}
\caption{}
\label{g6}
\end{center}
\end{figure} 
 
 If we collect many such particles in two boxes and then compress the contents of the boxes to form a pair of black holes, we expect they will be connected by a large ERB. In principle Alice and Bob can enter the two black holes and meet in the ERB.

\bn

\bf    
EXERCISE: Consider the experiment in figure \ref{g6} but with the addition of  ``monitor" qubits at the upper slit. The monitor qubits flip if a particle passes through the upper slit. 

Analyze the experiment from the \RSF viewpoint. Assuming the monitor qubits are also compressed to a monitor black hole what kind of ERB results?

\rm

\bn

There is nothing special about two slits. We can consider a three slit experiment.  A particle comes through the slits in a superposition of three states. Introducing three boxes the state can be written,
\be 
|100\ra +|010\ra + |001\ra
\label{Wtype}
\ee 
Evidently the three boxes share tripartite entanglement,
but not of the GHZ type. The state \ref{Wtype} is called the W-state, for what reason I do not know.
W-entanglement is different than GHZ. Each qubit is again entangled with the union of the other two, but less than maximally. Unlike the GHZ state, any two qubits are entangled although  not quite maximally.

By collecting such particles in three boxes and collapsing them a tripartite W-entangled system of black holes can be created. The W-entanglement should allow messages from any two parties to meet.

\bn

Understanding the geometry of tripartite and multipartite entangled black holes, as well as the physical interpretation of them is a challenge worth pursing.

\section{So What?}
What I have done in this lecture is trivial. I've taken some ordinary quantum phenomena and quantum protocols, and by invoking ER=EPR I've reinterpreted them in terms of the geometry of Einstein-Rosen bridges. No new phenomena were discovered  other than the correlation with what infalling observers see, and whether they can meet behind the horizons of the ERB. The interesting thing is that such a translation is at all possible.

\bn

 The current source of all wisdom, AdS/CFT, has provided tremendous inspiration and knowledge about quantum gravity, but it is not all there is. Why is it that in AdS/CFT we never have to talk about those questions that the \RSF  addresses? There is a  reason: the existence of an  asymptotic  boundary. The  theory is set up so that an outside ``uber-observer"  can manipulate the CFT, and make measurements on it, but the uber-observer is not part of the system. For the purposes of the uber-observer  the \CI (and the collapse of the wave function) is a sufficient framework.

Such an uber-observer makes things easy but unsatisfying. Sooner or later we will have to give up the security of an asymptotically cold boundary, and formulate a theory in which the universe is a highly interconnected network of entangled subsystems, with no preferred uber-observer. I expect that when this happens ER=EPR will take its place as one of the cornerstones of the new theory.

 What all of this suggests to me, and what I want to suggest to you, is that quantum mechanics and gravity are far more tightly related than we (or at least I) had ever imagined. The essential nonlocalities of quantum mechanics---the need for instantaneous communication in order to classically simulate entanglement---parallels the nonlocal potentialities of general relativity: ER=EPR.

\section*{Acknowledgements}

I would like to thank Hrant Gharibyan, Sandu Popescu, Ying Zhao, and  for discussions concerning some of the material in this lecture. I am also grateful to Grant Salton, Aron Wall, Brian Swingle, and Adam Brown for pointing out errors and oversights in the original version.

I would also like to thank the Institute for Advanced Study for hospitality.

This work was supported in
part by National Science Foundation grant 0756174 and by a grant from the John Templeton Foundation.
The opinions expressed in this publication are those of the author and do not necessarily
reflect the views of the John Templeton Foundation.

\end{document}